\documentclass[amsmath,amssymb,twocolumn,floatfix,superscriptaddress,preprintnumbers,nofootinbib,prd]{revtex4-2}

\input{header}

\usepackage{listings}
\usepackage{float}

\makeatletter
\def\l@subsection#1#2{}
\def\l@subsubsection#1#2{}
\makeatother

\begin{document}

\preprint{UCI-TR-2024-01}

\title{Simulating Heavy Neutral Leptons with General Couplings at Collider and Fixed Target Experiments}

\author{Jonathan L.~Feng}
\email{jlf@uci.edu}
\affiliation{Department of Physics and Astronomy, University of California, Irvine, CA 92697, USA}

\author{Alec Hewitt}
\email{ahewitt1@uci.edu}
\affiliation{Department of Physics and Astronomy, University of California, Irvine, CA 92697, USA}

\author{Felix Kling}
\email{felix.kling@desy.de}
\affiliation{Deutsches Elektronen-Synchrotron DESY, Notkestr.~85, 22607 Hamburg, Germany}

\author{Daniel La Rocco}
\email{laroccod@uci.edu}
\affiliation{Department of Physics and Astronomy, University of California, Irvine, CA 92697, USA}

\begin{abstract}
Heavy neutral leptons (HNLs) are motivated by attempts to explain neutrino masses and dark matter.  If their masses are in the MeV to several GeV range, HNLs are light enough to be copiously produced at collider and accelerator facilities, but also heavy enough to decay to visible particles on length scales that can be observed in particle detectors.  Previous studies evaluating the sensitivities of experiments have often focused on simple, but not particularly well-motivated, models in which the HNL mixes with only one active neutrino flavor.  In this work, we accurately simulate models for HNL masses between 100 MeV and 10 GeV and arbitrary couplings to $e$, $\mu$, and $\tau$ leptons. We include over 150 HNL production channels and over 100 HNL decay modes, including all of the processes that can be dominant in some region of the general parameter space.  The result is \texttt{HNLCalc}, a user-friendly, fast, and flexible library to compute the properties of HNL models. As examples, we implement \texttt{HNLCalc} to extend the \texttt{FORESEE} package to evaluate the prospects for HNL discovery at forward LHC experiments. We present sensitivity reaches for FASER and FASER2 in five benchmark scenarios with coupling ratios $|U_e|^2 : |U_\mu|^2 : |U_{\tau}|^2$ = 1:0:0, 0:1:0, 0:0:1, 0:1:1, and 1:1:1, where the latter two have not been studied previously. Comparing these to current constraints, we identify regions of parameter space with significant discovery prospects.  
\end{abstract}

\maketitle

\section{Introduction}
\label{sec:intro} 

The Standard Model (SM) is a remarkably successful theory of particle physics and includes all of the observed particles in nature. However, it does not explain all of the observed phenomena in nature, and hence it is incomplete. In particular, the SM does not accommodate the observed neutrino masses and mixings, and none of the particles it includes can be a significant fraction of dark matter.

Among the simplest ways to extend the SM is to introduce additional fermions that are uncharged under all SM gauge symmetries.  Such fermions, known as sterile or right-handed neutrinos, immediately open avenues for addressing the aforementioned problems of the SM.  For neutrino masses, the introduction of sterile neutrinos leads to the appearance of neutrino masses and mixings, as required by experimental observations.  For dark matter, the coupling of sterile neutrinos to the SM through Yukawa couplings is the unique way that a dark fermion, that is, a fermion with no SM interactions, can interact with the SM through a renormalizable coupling, making it an especially important example of beyond-the-SM (BSM) physics. 

Once sterile neutrinos are introduced, they generically mix with the SM neutrinos, and the resulting mass eigenstates are often referred to as heavy neutral leptons (HNLs). HNLs are mostly sterile, but their small SM neutrino components imply that they do interact with SM gauge bosons, which may lead to observable signals.  Which signals are possible depends heavily on the HNL mass.  Mass scales that have been discussed at length in the literature include $\sim \text{eV}$ masses, motivated by experimental anomalies; $\sim \text{keV}$ masses, motivated by cosmology and possible evidence for warm dark matter; and masses at the TeV scale and above, motivated by models of leptogenesis and the see-saw mechanism.  For reviews, see, for example, Refs.~\cite{Drewes:2013gca, Deppisch:2015qwa, Abazajian:2017tcc, Dasgupta:2021ies}.  

In this work, we will consider HNLs with masses in the MeV to GeV range. Such HNLs have attracted interest since at least the early 1970's;~\cite{Bjorken:1972am,Shrock:1974nd}; for a review that includes references to the early literature, see Ref.~\cite{Bryman:2019bjg}. In recent years, however, there has been a resurgence of interest in HNLs with masses in the MeV to GeV range. HNLs with such masses may be used to generate neutrino masses and mixings consistent with experimental measurements and simultaneously address many of the cosmological problems of the SM; see, for example, Ref.~\cite{Asaka:2005pn}.  More generally, from a purely phenomenological perspective, such HNLs are amenable to a wide variety of experimental probes, since they are light enough to be copiously produced at many collider and accelerator facilities, but also heavy enough to decay to visible particles on length scales that can be observed in particle detectors~\cite{Shrock:1978ft, Gribanov:2001vv, Atre:2009rg, Helo:2010cw, Drewes:2015iva, Bondarenko:2018ptm, Ballett:2019bgd, Coloma:2020lgy, Ovchynnikov:2023cry}.  For these reasons, HNLs have become a leading example of long-lived particles (LLPs), and they have helped motivate the growing world-wide research program in search of LLPs.   

To evaluate the prospects for HNL discovery, it is essential to have a user-friendly, fast, and flexible tool that can model HNL production and decay and also be used to estimate event rates in current and proposed experiments. The number of production channels and decay modes that may be important for HNLs dwarfs the corresponding number for dark photons and other well-known LLPs. Thankfully, a great deal of work has been done to identify and quantify the leading production and decay processes~\cite{Shrock:1980vy, Shrock:1980ct, Shrock:1981wq, Gribanov:2001vv, Gorbunov:2007ak, Atre:2009rg, Helo:2010cw, Ballett:2019bgd, Coloma:2020lgy}. Building on this work, in this study, we present the python library \texttt{HNLCalc}, which computes the properties of general HNL models.\footnote{\texttt{HNLCalc} is publicly available on GitHub at \github{https://github.com/laroccod/HNLCalc}.} It includes all of the potentially dominant processes and accurately describes the production and decay of HNLs with $\mathcal{O}$(GeV) masses and arbitrary couplings to the $e$, $\mu$, and $\tau$ leptons. 

HNLs may be produced in both fixed target and particle collider experiments. Fixed target experiments may produce many HNLs in the MeV to GeV range, and it is certainly of interest to predict the sensitivity of current and proposed experiments for general HNL searches.  At the same time, particle colliders, like the LHC, are also of interest, particularly in the forward direction, where event rates for such HNLs are greatly enhanced.  Our results may be used to model general HNL models and determine discovery prospects at both accelerator and collider experiments.

As examples, in this study, we consider the Forward Search Experiment (FASER)~\cite{Feng:2017uoz, FASER:2022hcn}, a current experiment that is purpose-built to search for LLPs in the far-forward region at the LHC, and FASER2, a future experiment to be housed in the proposed Forward Physics Facility (FPF)~\cite{Anchordoqui:2021ghd, Feng:2022inv} at the High-Luminosity LHC (HL-LHC), which will have an even greater discovery potential for LLPs.   For forward physics experiments at colliders, the FORward Experiment SEnsitivity Estimator (\texttt{FORESEE})~\cite{Kling:2021fwx} simulation package has become a very useful tool.  For a fixed proton-proton center-of-mass (COM) energy, \texttt{FORESEE} takes as input the forward hadron production rates given by MC generators, determines the resulting energy and angular distribution for various LLPs, and then calculates the signal rate in particle detectors.  \texttt{FORESEE} modules have been written to simulate a variety of LLPs, including dark photons, other gauge bosons, and also scalars, but, before this work, not HNLs. In this work, we extend \texttt{FORESEE} to simulate HNLs, using the HNL properties provided by the \texttt{HNLCalc} package.  We will focus on models with Majorana-like HNLs in which total lepton number $L$ is violated. 

To illustrate the flexibility of both \texttt{HNLCalc} and the \texttt{FORESEE} HNL module, we analyze the sensitivity reach for five benchmark models, where the ratios of HNL couplings are
\begin{align}
& \ \  |U_e|^2 : |U_{\mu}|^2 : |U_{\tau}|^2  \nonumber \\
\text{Benchmark 1}: & \quad \ 1 \ \ \, : \ \ \ 0 \ \ \, : \ \ 0 \nonumber \\    
\text{Benchmark 2}: & \quad \ 0 \ \ \, : \ \ \ 1 \ \ \, : \ \ 0 \nonumber \\  
\text{Benchmark 3}: & \quad \ 0 \ \ \, : \ \ \ 0 \ \ \, : \ \ 1 \nonumber \\   
\text{Benchmark 4}: & \quad \ 0 \ \ \, : \ \ \ 1 \ \ \, : \ \ 1 \nonumber \\   
\text{Benchmark 5}: & \quad \ 1 \ \ \, : \ \ \ 1 \ \ \, : \ \ 1 \ .  \label{eq:benchmarks} 
\end{align}
We will refer to these benchmarks using the convenient shorthand 100, 010, 100, 011, and 111.  The 100, 010, and 001 models are simple cases, each with only one nonzero coupling, and they are among the benchmarks typically considered, for example, by the Physics Beyond Colliders study group~\cite{Beacham:2019nyx}. The reach for FASER in these scenarios has been analyzed previously in Refs.~\cite{Kling:2018wct,Helo:2018qej,FASER:2018eoc}, and we reproduce these earlier results. The 011 and 111 models are less minimal, but have been proposed as more representative of models that explain the observed neutrino masses and mixings~\cite{Drewes:2022akb}.  The reach for FASER and FASER2 in the 011 and 111 models has not been determined previously, but is analyzed here rather easily, given the flexibility of the work described here.  We note that sensitivities for models with more than one non-zero coupling cannot be estimated simply by adding together event rates from the 100, 010, and 001 models, since all couplings enter the decay width, and so turning on a second or third coupling impacts both the event rates and the kinematic distributions of events mediated by the first coupling in highly non-trivial ways.  This interplay can only be included through the detailed simulation work described here.

This paper is organized as follows: In \cref{sec:model}, we discuss the HNL model and establish the notation we will use.  In \cref{sec:production,sec:decays} we discuss the many HNL production and decay processes, respectively, and we describe the detector and collider configuations we will consider in \cref{sec:FPF}. In \cref{sec:reach} we use all of these results to determine the sensitivity and discovery prospects for FASER and FASER2 in the five benchmark models discussed above.  We collect our conclusions in \cref{sec:conclusions}.  Detailed expressions for the production and decay branching fractions are contained in \cref{apx:production,apx:decay}, respectively.  In the course of this work, typos and errors in the existing literature were identified, and these are noted in the Appendices.

\section{Model}
\label{sec:model}

In this paper, we consider the SM, with its three left-handed active neutrinos, extended to include $n$ right-handed sterile neutrinos.  With these extra states, the SM Lagrangian can be supplemented by additional gauge-invariant terms, such as 
\begin{align}
   \mathcal{L} \supset &  -\sum_{\alpha i} y_{\alpha  i}  \, \overline{L}_\alpha \,\tilde{\phi} \, N^\prime_i - \sum_{ij} m_{ij} \, \overline{{N^\prime_i}^c} \, N^\prime_j \nonumber \\
& - \sum_{\alpha \beta} \frac{1}{M}  \lambda_{\alpha \beta} \overline{L}_{\alpha}\, \tilde{\phi}\, \tilde{\phi}^T \,L_{\beta}^c  + \text{h.c.} \ ,
\label{eq:lagrangian}
\end{align}
where the fields are the SM left-handed lepton doublets $L_{\alpha} = (\nu_{\alpha}, l_{\alpha})^T$; the right-handed sterile neutrinos $N^\prime_i$, where the prime distinguishes the gauge eigenstates from the unprimed mass eigenstates to be defined below; the SM Higgs doublet $\phi$; and their charge conjugates  $L^c = C \overline{L}^T$, ${N^\prime}^c = C \overline{N^\prime}^T$,  and $\tilde{\phi} = i\sigma_2 \phi^*$.  The fermion fields $\nu_{\alpha}$, $l_{\alpha}$, and $N^\prime_i$ are each 4-component Weyl spinors (e.g., $N^\prime_i=( 0, {N_i^\prime}_R )^T$); the index sums are over $\alpha, \beta = e, \mu, \tau$ and $i, j = 1, \ldots, n$; $y_{\alpha  i}$ and $\lambda_{\alpha \beta}$ are dimensionless (Yukawa) couplings; and $m_{ij}$ and $M$ are mass parameters. 

The terms in the first sum in \cref{eq:lagrangian} are the neutrino Yukawa couplings, the only gauge-invariant and renormalizable terms that can couple SM fields to gauge singlet fermions.  The terms in the second sum are right-handed Majorana neutrino mass terms, which break total lepton number $L$. Last, the terms in the third sum also break $L$, but do not involve the $N^\prime_i$ fields. These terms are non-renormalizable, but can be generated once heavy singlet neutrinos are integrated out, as, for example, in the see-saw mechanism.  Similar $L$-violating terms may also be present even at the renormalizable level if one introduces additional fields, such as a scalar field that is a triplet of SU(2). 

After electroweak symmetry breaking, when the neutral component of the Higgs field obtains a vacuum expectation value, the terms of \cref{eq:lagrangian} all contribute to neutrino masses.   In the basis $(L_\alpha, {N^\prime_i}^c)$, the most general $(3+ n, 3+ n)$ mass matrix is
\begin{equation}
   M^\nu = \begin{pmatrix}
        M_L & M_D \\
        M_D^T & M_R
    \end{pmatrix} ,
\end{equation} 
where $M_D$, $M_R$, and $M_L$ are the Dirac, right-handed Majorana, and left-handed Majorana masses generated by terms in the first, second, and third sums of \cref{eq:lagrangian}, respectively.  When the mass matrix $M^{\nu}$ is diagonalized, the resulting mass eigenstates are $3+n$ Majorana neutrinos. These include the three mostly-active neutrinos that have been observed experimentally, $\nu_1$, $\nu_2$, and $\nu_3$, and the $n$ mostly-singlet HNLs, which we denote $N_i$, where $i=1, \ldots, n$.

In this study, for simplicity, we consider models in which the phenomenology is dominated by a single HNL, which we denote $N$, while others either contribute subdominantly or are outside the mass range of interest. Note, however, that we do not assume that the $N$'s mixing is dominated by mixing with only one active neutrino. We will consider the $N$ field to have a mass in the MeV to GeV range, and so a wealth of existing observables, for example, the width of the $Z$ boson, constrain their mixings with the active neutrinos to be small.  Barring rather special scenarios, for example, where the HNLs are almost mass-degenerate and can oscillate into each other on relevant length scales~\cite{Tastet:2019nqj, Antel:2023hkf}, the signal rates for HNLs in models with two or more HNLs with significant mixings can therefore be determined simply by adding together the signal rates for each HNL considered separately. 

With these simplifications, then, the neutrino flavor eigenstates can be expressed in terms of the mass eigenstates as
\begin{equation}
    \nu_\alpha = \sum_{i = 1}^{3} V_{\alpha i} \,\nu_i + U_\alpha N^c ,
\end{equation}
where $V_{\alpha i }$ and $U_\alpha$ parameterize the active and sterile neutrino content, respectively.  The SM couplings of the electroweak gauge bosons to neutrino flavor eigenstates therefore induce couplings to $N$ proportional to $U_{\alpha}$.  The charged-current (CC) and neutral-current (NC) interaction terms are
\begin{equation}
\begin{split}
\mathcal{L}^{\text{CC}} &= - \frac{g}{\sqrt{2}}\sum_{\alpha} U^*_\alpha  W_\mu^+  \,\overline{N^c} \, \gamma^\mu l_\alpha + \text{h.c.} \\
\mathcal{L}^{\text{NC}} &=  - \frac{g}{2\cos\theta_W}\sum_{\alpha} U^*_\alpha Z_\mu  \overline{N^c}\, \gamma^\mu \nu_\alpha+ \text{h.c.}
    \label{eq:CCNC}
\end{split}
\end{equation} 
The production and decay rates for the HNL are completely determined by the mass $m_N$ and the couplings $U_\alpha$.  As noted in \cref{sec:intro}, in this study, we consider benchmark models with fixed coupling ratios $|U_e|^2$\,:\,$|U_\mu|^2$\,:\,$|U_\tau|^2$ defined in \cref{eq:benchmarks}.  The only remaining freedom, then, is the overall size of these couplings, which we parameterize by $\epsilon$, the sum of the $U_{\alpha}$ couplings in quadrature:
\begin{equation}
\label{eq:coupling}
    \epsilon^2 = |U_e|^2 + |U_\mu|^2 + |U_\tau|^2 \ .
\end{equation}
With these definitions, given a particular benchmark model, the HNL phenomenology is completely determined by the two parameters 
\begin{equation}
m_N, \epsilon \ .
\end{equation}
We will present all of our results below, including the sensitivity reaches of various experiments, as functions of one of these two parameters or in the $(m_N, \epsilon)$ plane.

\section{HNL Production}
\label{sec:production}

At the LHC, high-energy proton-proton collisions produce hadrons and leptons that can decay to HNLs through the CC and NC interactions of \cref{eq:CCNC}.  In the mass range of interest here, HNLs are primarily produced in the decays of mesons and tau leptons. The production of these parent particles at the LHC is modeled using a variety of generators, which will be discussed below in \cref{subsec: HNL production Rates}.

\begin{table*}[tbp]
    \centering
    \begin{tabular}{ c | c |c c c c c c}
\hline\hline
\multicolumn{8}{c}{\centering HNL Production Processes }\\
\hline\hline
$P \to l N$ &  \cref{fig:feynman_prod} (a) 
& $\pi^+ \to l^+ N$                 & $K^+ \to l^+ N$  
& $D^+ \to l^+ N$                   & $D_s^+ \to l^+ N$ 
& $B^+ \to l^+ N$                   & $B_c^+ \to l^+ N$ 
\\
\hline
$P \to P^\prime l N$ & \cref{fig:feynman_prod} (b) 
& $K^+ \to \pi^0 l^+ N$             &  $K_S \to \pi^{+} l^{-} N$ 
& $K_L \to \pi^{+} l^{-} N$         & $D^0 \to K^- l^+ N$
& $\bar{D}^0 \to \pi^+ l^- N$       & $D^+ \to \pi^0 l^+ N$\\
&&$D^+ \to \eta l^+ N$              & $D^+ \to \eta^\prime l^+ N$
& $D^+ \to \bar{K}^0 l^+ N$         & $D_s^+ \to \bar{K^0} l^+ N$
& $D_s^+ \to \eta l^+ N$            & $D_s^+ \to \eta^\prime l^+ N$\\
&&$B^+ \to \pi^0 l^+ N$             & $B^+ \to \eta l^+ N$
& $B^+ \to \eta^\prime l^+ N$       & $B^+ \to \bar{D}^0 l^+ N$ 
& $B^0 \to \pi^- l^+ N$             & $B^0 \to D^- l^+ N$\\
&&$B^0_s \to K^- l^+ N$             & $B^0_s \to D^-_s l^+ N$ 
& $B^+_c \to D^0 l^+ N$             & $B^+_c \to \eta_c l^+ N$
& $B^+_c \to B^0 l^+ N$             & $B^+_c \to B^0_s l^+ N$\\ 
\hline
$P \to V l N$ & \cref{fig:feynman_prod} (b)
& $D^0 \to \rho^- l^+ N$            & $D^0 \to K^{*-} l^+ N$ 
& $D^+ \to \rho^0 l^+ N$            & $D^+ \to \omega l^+ N$ 
& $D^+ \to \bar{K}^{*0} l^+ N$      & $D^+_s \to K^{*0} l^+ N$\\
&&$D^+_s \to \phi l^+ N$            & $B^+ \to \rho^0 l^+ N$
& $B^+ \to \omega l^+ N$            & $B^+ \to \bar{D}^{*0} l^+ N$
& $B^0 \to \rho^- l^+ N$            & $B^0 \to D^{*-} l^+ N$\\
&&$B^0_s \to K^{*-} l^+ N$          & $B^0_s \to D^{*-}_s l^+ N$
& $B_c^+ \to D^{*0} l^+ N$          & $B_c^+ \to J/\psi \; l^+ N$ 
& $B^+_c \to B^{*0} l^+ N$          & $B^+_c \to B^{*0}_s l^+ N$ \\
\hline
$\tau \to H N$ & \cref{fig:feynman_prod} (c)
& $\tau^+ \to \pi^+ N$              & $\tau^+ \to K^+ N$
& $\tau^+ \to \rho^+ N$             & $\tau^+ \to K^{*+} N$ &&\\
\hline
$\tau^+ \to l^+ \nu N$ & \cref{fig:feynman_prod} (d,e)
& $\tau^+ \to l^+ \bar{\nu}_\tau N$ & $\tau^+ \to l^+ \nu_l N$ 
&&&& \\ 
\hline \hline
\end{tabular}
\caption{HNL production processes included in \texttt{HNLCalc}. The processes are ordered by increasing parent particle mass; $P$, $V$, and $H$ denote pseudoscalar mesons, vector mesons, and hadrons, respectively; $l=e,\mu,\tau$; and $N$ is the HNL.  Charge-conjugate processes are also implemented, but are not explicitly listed here.  Representative Feynman diagrams for these processes are shown in \cref{fig:feynman_prod}.
}
\label{tab:production modes}
\end{table*}

\begin{figure*}[tbp]
\centering
\begin{tabular}{@{}c@{}}
   \\ \\
   \includegraphics[width=0.30\textwidth]{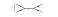} 
   \\ \\ [\abovecaptionskip]
   \small{(a) 2-body decay $P \to l N$}
\end{tabular} \hspace{.5cm}
\begin{tabular}{@{}c@{}}
   \includegraphics[width=0.33\textwidth]{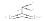} 
   \\ [\abovecaptionskip]
   \small{(b) 3-body decays $P \to P^\prime l N$ and $P \to V l N$}
\end{tabular} \\ \vspace{0.5cm}
\begin{tabular}{@{}c@{}}
   \includegraphics[width=0.28\textwidth]{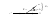}
   \\ [\abovecaptionskip]
   \small{(c) 2-body decay $\tau \to H N$}
\end{tabular} \hspace{.5cm}
\begin{tabular}{@{}c@{}}
   \includegraphics[width=0.26\textwidth]{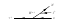}
   \\ [\abovecaptionskip]
   \small{(d) 3-body decay $\tau^+ \to l^+ \bar{\nu} N$}
\end{tabular} \hspace{.5cm}
\begin{tabular}{@{}c@{}}
   \includegraphics[width=0.26\textwidth]{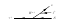}
   \\ [\abovecaptionskip]
   \small{(e) 3-body decay $\tau^+ \to l^+ \nu N$}
\end{tabular} \hspace{.5cm}
\caption{Representative Feynman diagrams for the HNL production processes listed in \cref{tab:production modes}. The subscripts $m$ and $n$ are generation indices.}
\label{fig:feynman_prod}
\end{figure*}

The complete list of all HNL production processes included in this work is given in \cref{tab:production modes}.  They include both 2-body and 3-body decays, and are divided into six categories: $P \to l N$, $P \to P^\prime l N$, $P \to V l N$, $\tau \to H N$, $\tau^+ \to l^+ \bar{\nu}_\tau N$, and $\tau^+ \to l^+ \nu_\ell N$, where $P$, $V$, and $H$ represent pseudoscalar mesons, vector mesons, and hadrons, respectively. Representative quark-level Feynman diagrams for these processes are given in \cref{fig:feynman_prod}, and expressions for the corresponding branching fractions are given in \cref{apx:production}. 

The HNL production processes of \cref{tab:production modes} include all of the possibly leading contributions.  Note that all of the parent hadrons are pseudoscalar mesons. Vector meson decays do not typically produce HNLs with significant branching fractions, because the decays to HNLs compete with decays mediated by the strong interactions and so are highly suppressed.  HNL production in baryon decays is also not included, since they are subdominant.

\begin{figure*}[tbp]
\centering
\includegraphics[width=0.48\linewidth]{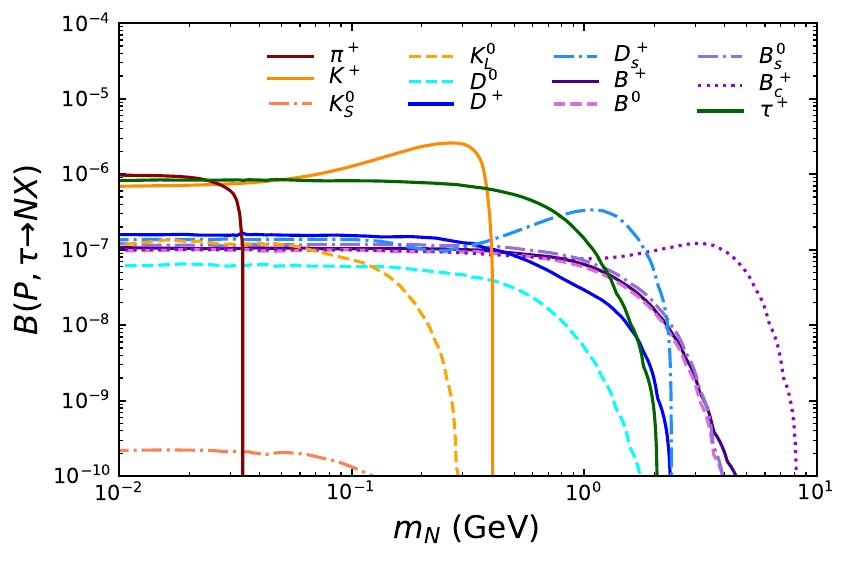} 
\includegraphics[width=0.48\linewidth]{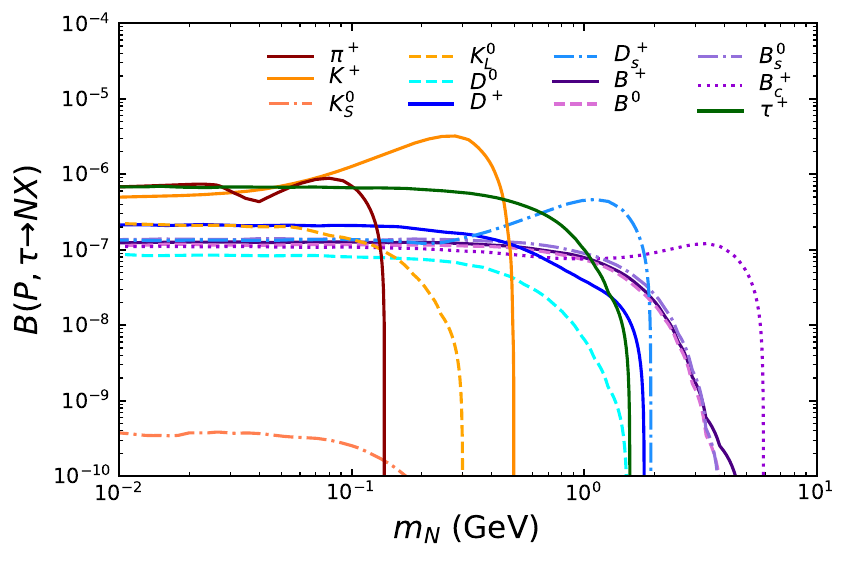}
\caption{Branching fractions $B(P, \tau \to NX)$ as functions of the HNL mass $m_N$ for the 011 (left) and 111 (right) benchmarks and $\epsilon = 10^{-3}$.  $P$ represents pseudoscalar mesons, and $NX$ represents any final state containing the HNL $N$. }
\label{fig:grouped}
\end{figure*}

The branching fractions for meson and tau lepton decays into HNL’s are presented in \cref{fig:grouped}, as functions of HNL mass. For illustrative purposes, we show results for the 011 and 111 benchmark models with coupling parameter $\epsilon = 10^{-3}$. For each benchmark model there are as many as 150 HNL production processes.  Rather than show branching fractions for each of these modes, in \cref{fig:grouped} we show the total BSM branching fraction for each parent meson, which includes final states with all possible hadrons and lepton flavors.  

Because the total branching fractions include many modes, distinctive aspects of individual modes are not always apparent.  However, we note a few general features:
\begin{itemize}[itemsep=0.03cm,topsep=0.15cm,leftmargin=1.5em]
\item The branching fractions are typically larger for the longer-lived mesons, where the competing SM decay modes having smaller widths.  This implies that the BSM branching fractions are typically larger for the lighter parent mesons.  
\item Of course, the branching fractions vanish when the decay modes become kinematically inaccessible. For example, $B(\pi^+ \to NX)$ vanishes at $m_N = m_{\pi^+} - m_{\mu} \simeq 34~\mev$ for the 011 benchmark and at $m_N = m_{\pi^+} - m_{e} \simeq 139~\mev$ for the 111 benchmark. 
\item At the same time, although for many modes the branching fraction drops as $m_N$ approaches the kinematic threshold, this is not always the case. For parent mesons where the dominant production mode is chirality suppressed, for example, for $K^+$ mesons, where the dominant HNL decay mode is $K^+ \to l^+ N$, the branching fraction vanishes for $m_N = 0$, and grows as $m_N$ increases.  In these cases, the branching fraction can be large and even maximal very near threshold.
\end{itemize}
These facts, coupled with the fact that more light mesons than heavy mesons are produced at the LHC, imply that typically the dominant production mechanism for HNLs is decays of the lightest parent mesons for which the decay is kinematically allowed.  

In \cref{fig:grouped}, we have fixed $\epsilon = 10^{-3}$, a value that is very roughly at the limit of current constraints.  We see that branching fractions of $10^{-7}$ to $10^{-6}$ are allowed.  Given the enormous flux of far-forward mesons at the LHC, this implies that the flux of highly collimated far-forward HNLs can be significant.  Of course, to be detected, the HNLs must decay in the detector to a visible final state.  

\section{HNL Decays}
\label{sec:decays}

\begin{figure*}[tbp]
\centering
\begin{tabular}{@{}c@{}}
   \includegraphics[width=0.26\textwidth]{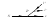}
   \\ [\abovecaptionskip]
   \small{(a) NC-mediated $N\to \nu l^+ l^-$}
\end{tabular} \hspace{.5cm}
\begin{tabular}{@{}c@{}}
   \includegraphics[width=0.26\textwidth]{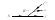}
   \\ [\abovecaptionskip]
   \small{(b) CC-mediated $N\to \nu {l^\prime}^+ l^-$}
\end{tabular}\hspace{.5cm}
\begin{tabular}{@{}c@{}}
   \includegraphics[width=0.26\textwidth]{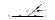}
   \\ [\abovecaptionskip]
   \small{(c) NC-mediated $N\to \nu\bar{\nu}\nu$}
\end{tabular}\\ \vspace{0.5cm}
\begin{tabular}{@{}c@{}}
   \includegraphics[width=0.32\textwidth]{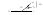} 
   \\ [\abovecaptionskip]
   \small{(d) NC-mediated $N\to \nu H^0$}
\end{tabular}\hspace{1cm}
\begin{tabular}{@{}c@{}}
   \includegraphics[width=0.32\textwidth]{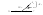}
   \\ [\abovecaptionskip]
   \small{(e) CC-mediated $N\to l^- H^+$}
\end{tabular}
\caption{Representative Feynman diagrams for HNL decays. The subscripts $m,n$ are generation indices. }
\label{fig:feynman_decay}
\end{figure*}

Once produced, HNLs decay to SM final states through the CC and NC interactions of \cref{eq:CCNC}. The complete list of HNL decay modes included in this work is given in \cref{tab:decay modes}. We consider purely leptonic 3-body decays, semi-leptonic 2-body decays, and semi-leptonic 3-body decays. Representative Feynman diagrams for each category are shown in \cref{fig:feynman_decay}. The HNL decay branching fractions are computed using formulae derived in Refs.~\cite{Coloma:2020lgy,Bondarenko:2018ptm}, and the expressions for the branching fractions are given in \cref{apx:decay}. 

\setlength{\tabcolsep}{8pt} 
\begin{table*}[tbp]
    \centering
    \begin{tabular}{c | c |c c c c c c} 
    \hline\hline
    \multicolumn{8}{c}{HNL Decay Modes}\\
    \hline\hline
    $\nu\, l^+ l^-$ & \cref{fig:feynman_decay} (a) 
    & $\nu_l  e^+ e^-$ & $\nu_l  \mu^+ \mu^-$ & $\nu_l  \tau^+ \tau^-$   &&&  \\
    \hline 
    $l^\pm \, \nu_{l'} l'^\mp $ & \cref{fig:feynman_decay} (b) 
    & $l^\pm \nu_e e^\mp$ & $l^\pm \nu_\mu \mu^\mp$ & $l^\pm \nu_\tau \tau^\mp$ &&& \\
    \hline
    $\nu_l \, \overline{\nu} \nu$ &  \cref{fig:feynman_decay} (c) 
    & $ \nu_l  \bar{\nu}_e \nu_e$ & $\nu_l \bar{\nu}_\mu \nu_\mu$ & $\nu_l \bar{\nu}_\tau \nu_\tau$ &&& \\ 
    \hline
    $\nu_l \, H^0$  & \cref{fig:feynman_decay} (d) &
    $\nu_l  \pi^0$ & $\nu_l  \eta$ & $\nu_l \eta'$ & $\nu_l \rho^0$ & $\nu_l \omega$ & $\nu_l \phi$ \\ 
    \hline
    $l^\pm \, H^\mp$  & \cref{fig:feynman_decay} (e) &
    $l^\pm  \pi^\mp$ & $l^\pm  K^\mp$ & $l^\pm  D^\mp$ & $l^\pm  D_s^\mp$ & $l^\pm  \rho^\mp$ & $l^\pm  K^{*\mp}$ \\
    \hline \hline 
    $\nu_l \,q \overline{q} $  & \cref{fig:feynman_decay} (d) 
    & $\nu_l u \overline{u}$ & $\nu_l d \overline{d}$ & $\nu_l s \overline{s}$ & $\nu_l c \overline{c}$ & $\nu_l b \overline{b}$ \\
    \hline
    $l^\pm \,u \overline{d}^\prime $   & \cref{fig:feynman_decay} (e) &
    $l^- u \overline{d}$ & $l^-u \overline{s}$ & $l^- u \overline{b}$ & $l^+ \overline{u} d$ & $l^+  \overline{u} s$ & $l^+  \overline{u} b$ \\ 
    && $l^- c \overline{d}$ & $l^- c \overline{s}$ & $l^- c \overline{b}$ & $l^+ \overline{c} d$ & $l^+  \overline{c} s$ & $l^+  \overline{c} b$ \\ 
    \hline \hline 
    \end{tabular}
\caption{HNL decay modes included in \texttt{HNLCalc}, where $H$ denotes hadrons. Representative Feynman diagrams for these modes are shown in \cref{fig:feynman_decay}. Quark level decays, $\nu q \bar{q}$ and $lu\bar{d}$, are used when computing the total hadronic width when  $m_N >1.0$ GeV. }
\label{tab:decay modes}
\end{table*}

To properly simulate the response of the detector to an HNL decay, it is important to know and accurately represent its final states. For example, it can make a big difference experimentally whether an HNL decays to states with charged tracks, e.g., $N\to \nu \rho \to \nu \pi^+\pi^-$, or to states with only final state photons, e.g., $N\to \nu\pi^0 \to \nu\gamma\gamma$. For this reason, it is generally insufficient to specify HNL decays into quarks, as it would rely on hadronization tools to obtain hadronic final states. These tools are known not to work well when the invariant mass of the hadronic final state is close to or below the QCD confinement scale, because, for example, this treatment fails to model hadronic resonances and kinematic thresholds. More generally, in this regime, the factorization theorem loses validity, meaning that one cannot factorize the HNL decay into quarks and use quark hadronization into hadrons anymore. On the other hand, for $m_N \agt 1.0$ GeV, decays into single mesons are insufficient for computing the total HNL decay width, since in this region multi-meson decays become important. 

To best model hadronic decays at all scales, while also accurately computing the HNL lifetime, we follow the approach taken in Refs.~\cite{Coloma:2020lgy,Bondarenko:2018ptm}. For $m_N < 1.0~\gev$, decays into single mesons, $H = \pi^{\pm,0},K^\pm, \rho ^{\pm,0},\omega, K^{* \pm},\eta,\eta' $, are calculated using their respective decay constants, $f_H$, and these are used to obtain the total hadronic decay width.   For $m_N \ge 1.0~\gev$, the total hadronic width is instead calculated by summing up HNL decay widths into the quark-level final states $lq \bar{q}'$ and $\nu q \bar{q}$.  A QCD loop correction to account for hadronization is applied. This estimated correction is obtained from the known corrections up to $\mathcal{O}(\alpha_s^3)$ in hadronic $\tau$ decay; see \cref{apx:decay} for its explicit form. 

Included in our single meson decays are decays into $\rho$ mesons. 
It has been shown in Ref.~\cite{Bondarenko:2018ptm} that the dominant source of two pion final states are decays into $\rho$ mesons followed by $\rho \to \pi \pi$.  We therefore take the difference between the quark-level decay width and the single-meson total width as an estimate of the decay width to final states with $\geq 3$ hadrons. Additionally, to avoid overestimating the contribution of $\nu s s$ to the hadronic width below the $2m_K$ kinematic threshold, a phase space suppression factor $\sqrt{1-4m_K^2/m_N^2}$ is applied to this decay. This same approach is also applied to the decays $\tau ud$ and $ \tau us$, with kinematic thresholds at $m_\tau+2m_\pi$ and $m_\tau + m_\pi + m_K$, respectively. 

\begin{figure*}[tbp]
\includegraphics[width=\linewidth]{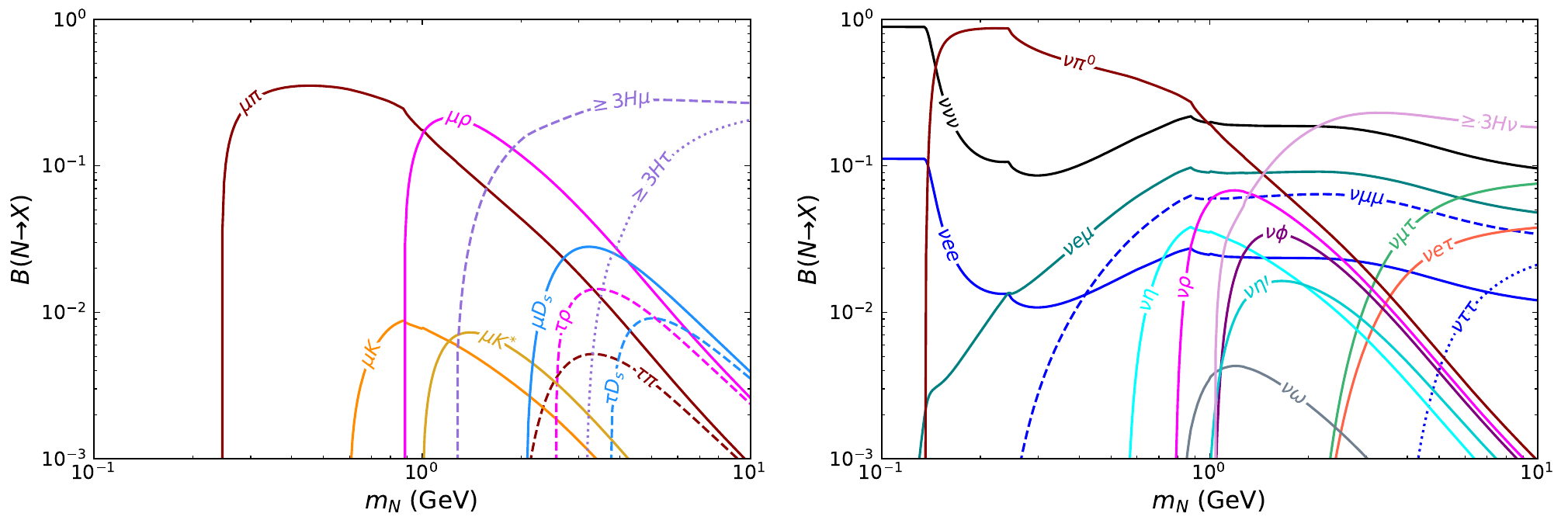} 
\caption{HNL branching fractions in the 011 benchmark model for the dominant decays via CC (left) and NC (right) interactions as functions of $m_N$. For $m_N \alt 1.0~\gev$, the decays are primarily to pions, while for $m_N \agt 1.0~\gev$, the decays are dominated by decays to 3 or more hadrons. Decays to $D$ mesons are subdominant, with $B(N \to lD) < 0.1 \%$.  }
\label{fig:decay br 011}
\end{figure*}

\begin{figure*}[tbp]
\includegraphics[width=\linewidth]{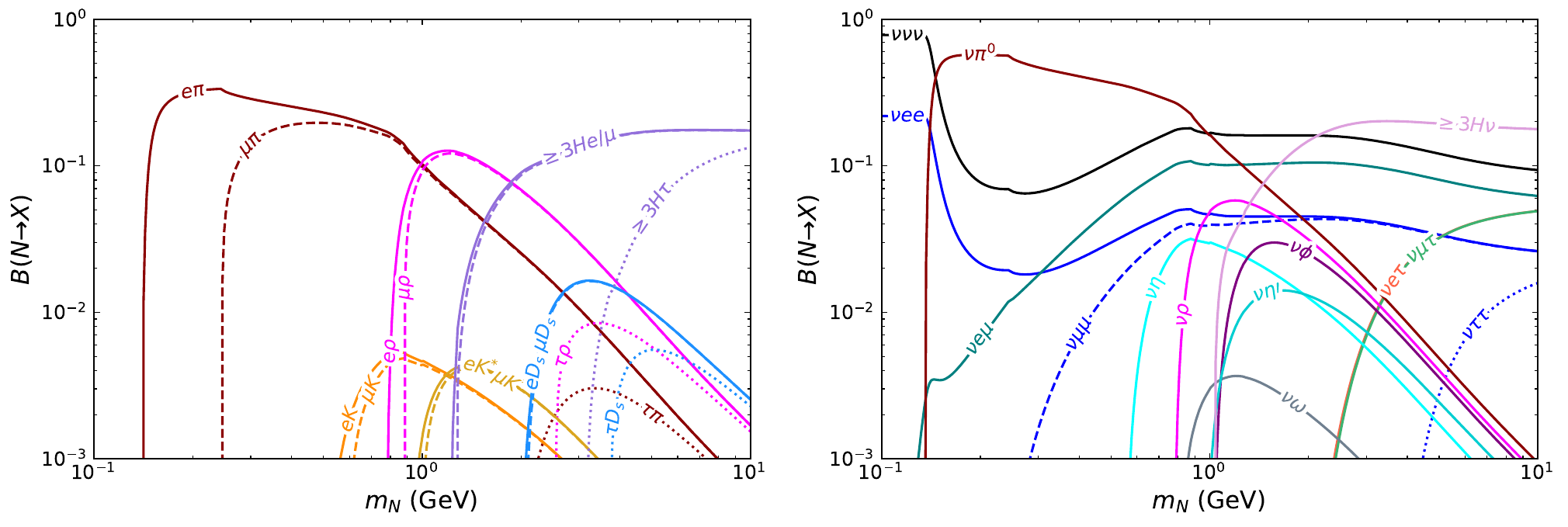} 
\caption{As in \cref{fig:decay br 011}, but for the 111 benchmark model. }
\label{fig:decay br 111}
\end{figure*}

\begin{figure}[tbp]
\includegraphics[width=0.95\linewidth]{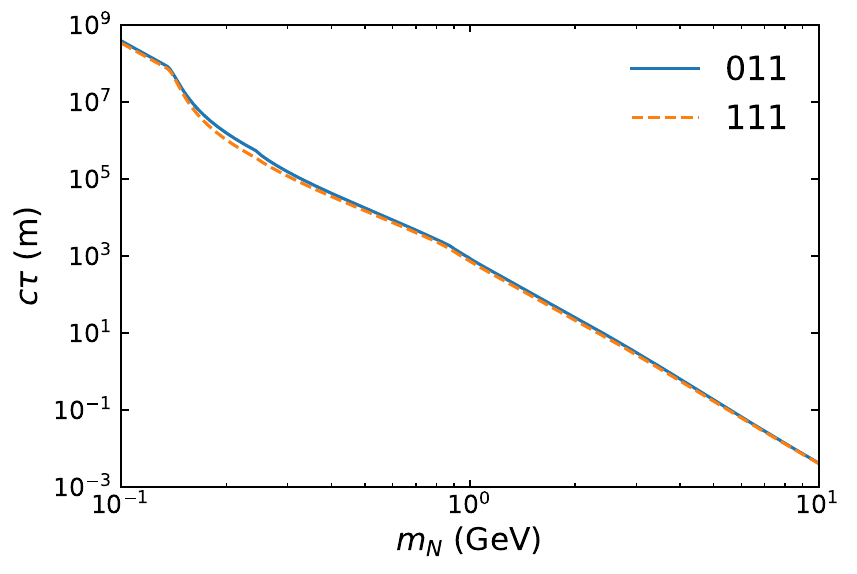} 
\caption{Lifetimes for the 011 and 111 benchmark models with $\epsilon = 10^{-3}$. A visible kink in the lifetime can be observed at $m_N\approx m_\pi$, where the decay $N\to \nu\pi^0$ becomes kinematically accessible.}
\label{fig:lifetime}
\end{figure}

In \cref{fig:decay br 011,fig:decay br 111}, we plot the branching fractions for all relevant modes in the 011 and 111 benchmarks. Additionally, the HNL lifetimes are given in \cref{fig:lifetime}.   For $m_N<m_{\pi}$, HNL decays are dominated by NC decays into the 3-body final states $\nu \bar{\nu} \nu$ and $\nu e^+e^-$. The invisible decay mode dominates, and the visible branching fraction is only approximately 10\% and 20\% in the 011 and 111 benchmark models, respectively. However, for $m_N > m_{\pi}$, the 2-body decays $N\to l^\pm \pi^\mp$ and $N \to \nu \pi^0$ become kinematically accessible, resulting in a sharp drop in the lifetime, as seen in \cref{fig:lifetime}. For masses $m_N > m_{\pi}$, the hadronic decay modes become dominant, and the invisible decay branching fraction is below 20\% in both scenarios. 

It is important to note that in this work we have neglected the effects of spin correlations between production and decay.  These do not change the HNL lifetimes and event rates, but they can impact the kinematic distributions of the final state particles. In the case of Majorana HNLs, where lepton-number-conserving and lepton-number-violating processes, for example, $B_c\to\mu^+\mu^+\tau^- \nu$ and $B_c\to\mu^+\mu^+\tau^- \bar{\nu}$, are both possible, interesting effects in kinematic distributions have the potential to differentiate between the two final states, which are otherwise experimentally identical; see, for example, Ref.~\cite{Vasquez:2023glo}. 

\section{HNLs at FASER and FASER2}
\label{sec:FPF}

In this section, we describe the modeling of meson and $\tau$ production at the LHC and the resulting flux of HNLs at FASER~\cite{Feng:2017uoz} and FASER2~\cite{Feng:2022inv} during Run 3 and the HL-LHC era.  Given the modeling of production and decay of HNLs in \cref{sec:production,sec:decays}, respectively, with the use of \texttt{HNLCalc}, the \texttt{FORESEE} simulation framework can determine the sensitivity for HNL searches with forward detectors at the LHC.

\subsection{Collider and Detector Setup}

FASER is located in a tunnel $L = 480\,\m$ downstream from the ATLAS interaction point (IP). The FASER decay volume is a cylinder with a radius of $R =10\,\cm$ and a length of $\Delta=1.5\,\m$ along the beam collision axis. The decay volume extends to angles $\theta \simeq 0.2~\text{mrad}$ from the beamline and pseudorapidities $\eta \agt 9.2$.  We will consider the reach of FASER during Run~3 of the LHC with an integrated luminosity of $250~\ifb$.  We will also consider the scenario in which the FASER detector runs throughout the HL-LHC era with an integrated luminosity of $3~\text{ab}^{-1}$.  

In addition, we consider FASER2, one of the experiments proposed for the future FPF~\cite{Anchordoqui:2021ghd, Feng:2022inv}. FASER2 will be positioned roughly $L=650\,\m$ away from the ATLAS IP.  Its decay volume has length $\Delta=10\,\m$ and a rectangular cross-section with dimensions $3\,\m \times 1\,\m$~\cite{osti_1972463}. The FASER2 decay volume extends to angles $\theta \simeq 2.4~\text{mrad}$ from the beamline and pseudorapidities $\eta \agt 6.7$. 

During Run 3, the LHC has operated with a $pp$ COM energy of $\sqrt{s}=13.6~\tev$.  For the HL-LHC era, the COM energy is expected to be increased to 14 TeV.  We have found negligible differences in sensitivities from this change in COM energy, and for simplicity, we assume $\sqrt{s}=14~\tev$ for all results derived here.  The detector configurations are summarized in \cref{tab:geom}.

\setlength{\tabcolsep}{2pt} 
\begin{table}[tbp]
    \centering
    \begin{tabular}{l |c |c |c |c}
    \hline \hline
    & $L$ & $\Delta$ & Geometry & $\mathcal{L}$\\
    \hline\hline
    FASER (Run 3) & $480\,\m$ & $1.5\,\m$ & Cyl.~$R = 10\,\cm$ & $250\,\text{fb}^{-1}$  \\ \hline
    FASER (HL-LHC) & $480\,\m$ & $1.5\,\m$ & Cyl.~$R = 10\,\cm$ & $3\,\text{ab}^{-1}$ \\ \hline
    FASER2 (HL-LHC) & $650\,\m$ & $10\,\m$ & Rect.~$3\,\m \times 1\,\m$& $3\,\text{ab}^{-1}$  \\ 
    \hline\hline
\end{tabular}
\caption{Parameters for the three experimental configurations considered: FASER (Run 3), FASER (HL-LHC), and FASER2 (HL-LHC). $L$ is the distance from the interaction point to the front of the detector, $\Delta$ is the length of the detector along the beam axis, Geometry specifies the cross sectional area (cylindrical for FASER, rectangular for FASER2), and $\mathcal{L}$ is the integrated luminosity.  Both FASER and FASER2 are assumed to be centered on the beam collision axis.}
\label{tab:geom}
\end{table}

\subsection{HNL Production Rates at the LHC}
\label{subsec: HNL production Rates}

HNL production rates at the LHC are obtained in \texttt{FORESEE} through the Monte-Carlo (MC) sampling of the decays of parent particles~\cite{Kling:2021fwx}. The spectra for parent pions and kaons are obtained using the dedicated hadronic interaction model \texttt{EPOS LHC}~\cite{Pierog:2013ria}. For charm and bottom hadrons, we use the spectra obtained from \texttt{POWHEG}~\cite{Alioli:2010xd} matched with \texttt{Pythia}~\cite{Bierlich:2022pfr}, as presented in Ref.~\cite{Buonocore:2023kna}. Tau leptons are produced primarily in charm and bottom hadron decay, so we follow the same prescription there.  The primary source of uncertainty on the HNL flux originates from the modeling of hadron production.  To estimate this uncertainty, we follow the prescription of Ref.~\cite{FASER:2024ykc} and consider the range of predictions from \texttt{EPOS-LHC}, \texttt{QGSJET~2.04}~\cite{Ostapchenko:2010vb}, and \texttt{Sibyll~2.3d}~\cite{Riehn:2019jet} to model the uncertainty for light hadron production, and we take account of scale variations~\cite{Buonocore:2023kna} to model the flux uncertainty for heavy hadron production.
 
The production rates for the 011 and 111 benchmarks at the LHC are shown in \cref{fig:production rates,fig:production rates FASER2}. These figures show the total flux of HNLs produced in the direction of FASER and FASER2 respectively. For FASER, \cref{fig:production rates} shows the rate for production with an angle $\theta \leq 0.1\,\text{m} / 480 \,\text{m} = 0.2\,\text{mrad}$ relative to the beam collision axis, and for FASER2, \cref{fig:production rates FASER2} shows the rate for the production of parent particles that pass through FASER2's $3~\text{m} \times 1~\text{m}$ transverse area. As anticipated in \cref{sec:production}, the dominant production rate is typically from the lightest meson for which the decay is kinematically accessible. For the 111 scenario, the decay $\pi \to e N$ is possible, and so this is the dominant production process for very light $N$ with masses $m_N \sim 100~\mev$.  For the 011 scenario, this decay is not allowed, and pion decays are never dominant for $m_N \agt 100~\mev$.  For larger $m_N$, the results for the 011 and 111 scenarios are similar, with the production rates cutting off as the kaon, tau lepton, $D$ meson, and $B$ meson kinematic thresholds are passed.

\begin{figure*}[tbp]
\centering
\includegraphics[width=0.48\linewidth]{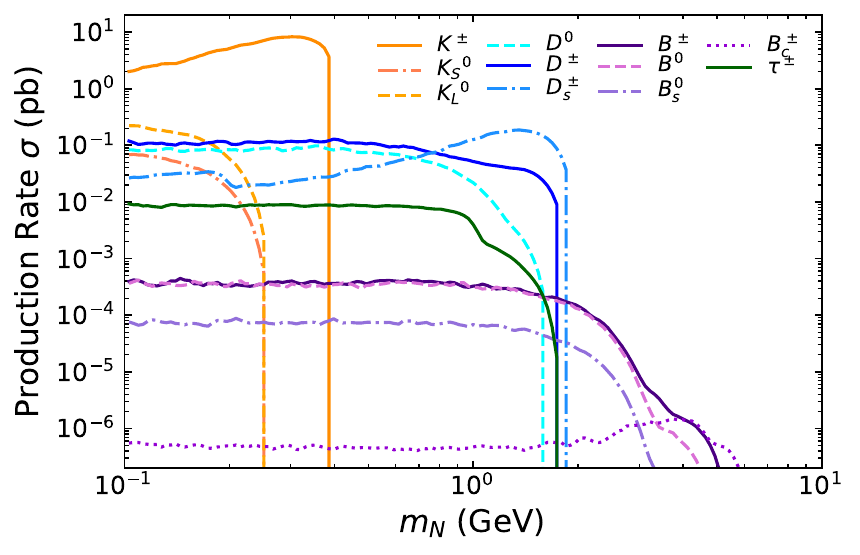} 
\includegraphics[width=0.48\linewidth]{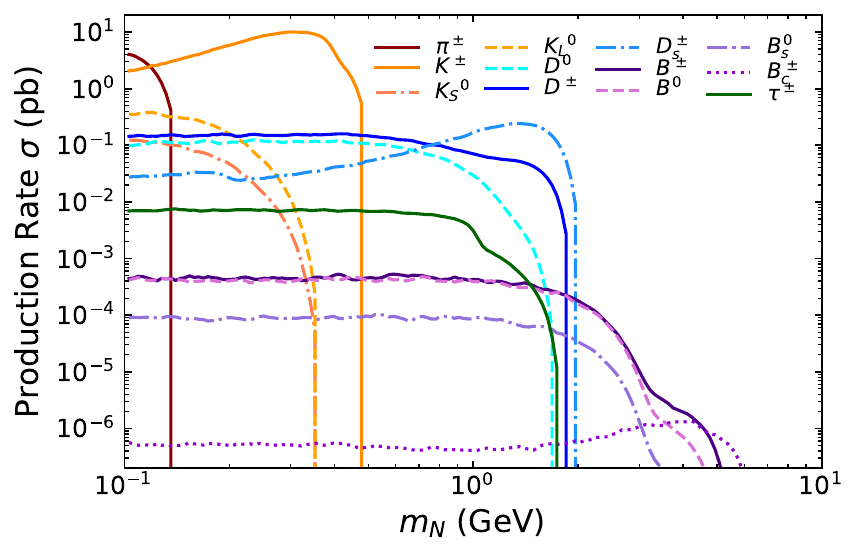}
\caption{Production rate of HNLs in the direction of FASER with $\theta< 0.2~\text{mrad}$. The rates are grouped by parent particle and are obtained by MC integration in \texttt{FORESEE} for the 011 (left) and 111 (right) benchmarks, with a coupling of $\epsilon = 10^{-3}$. }
\label{fig:production rates}
\end{figure*}

\begin{figure*}[tbp]
\centering
\includegraphics[width=0.48\linewidth]{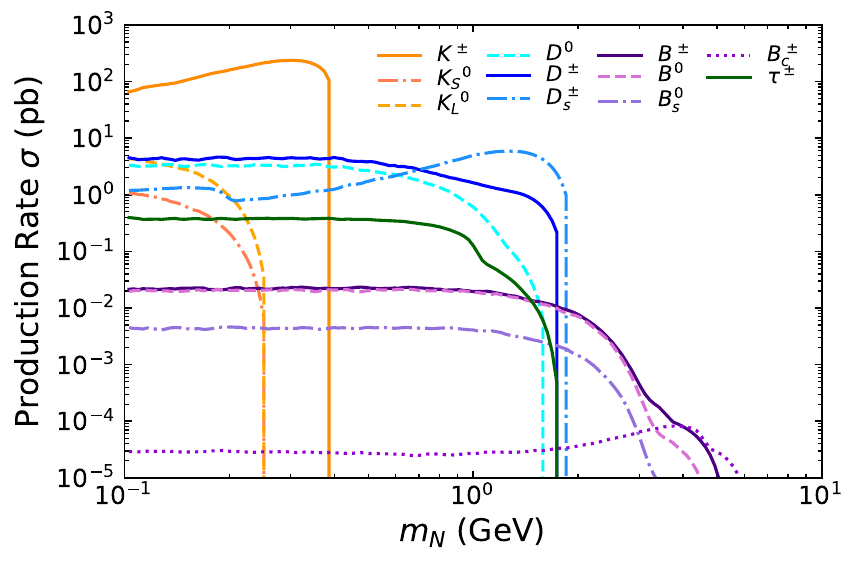}
\includegraphics[width=0.48\linewidth]{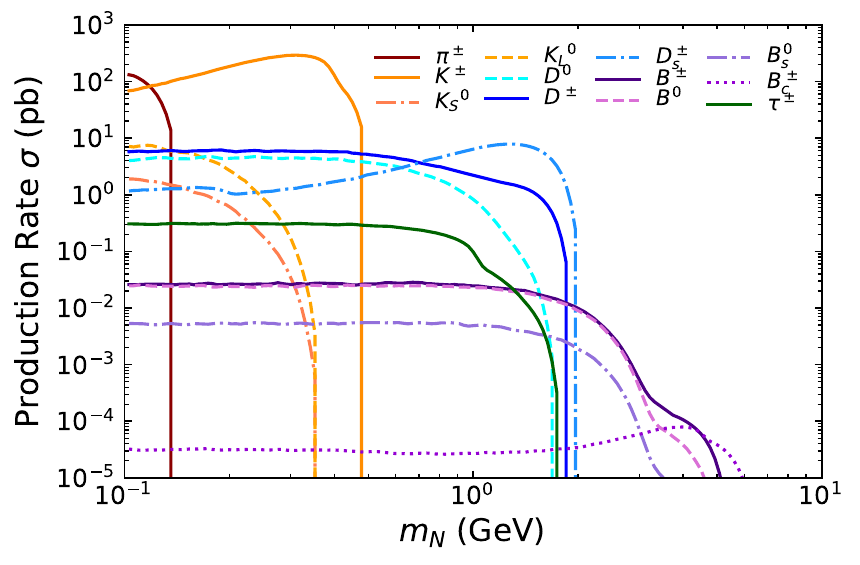}
\caption{Production rate of HNLs in the direction of FASER2, that is, that pass through FASER2's transverse area of $3~\text{m} \times 1~\text{m}$.  The rates are grouped by parent particle and are obtained by MC  integration in \texttt{FORESEE} for the 011 (left) and 111 (right) benchmarks, with a coupling of $\epsilon = 10^{-3}$. 
}
\label{fig:production rates FASER2}
\end{figure*}

\subsection{Signal and Background Considerations}
\label{subsec:backgrounds}

Depending on their mass, HNLs can decay into a variety of final states shown in \cref{fig:decay br 011,fig:decay br 111}. The dominant decay channels either consist of two charged particles, such as the decays $N \to \nu l^+ l^-$ and $N \to l^+ \pi^-$, or photons, especially from $N \to \nu \pi^0 \to \nu \gamma \gamma$. The corresponding experimental signatures in the FASER detector would consist of highly energetic charged particles or photons that emerge from the decay volume. The charged particle signal would leave high-momentum tracks in the FASER spectrometer, that are consistent with a single vertex inside the decay volume and whose combined momentum points back to the IP. If the final state contains electrons, they would additionally leave a sizbale energy deposit in the calorimeter. The multi-photon signal would leave a characteristic signal in the preshower and deposit a large amount of energy in the calorimeter. In both cases, the signal would not trigger the front veto.

The potential backgrounds for long-lived particle searches at FASER are induced by either high-energy muons or neutrinos coming from the direction of the IP. However, muons are efficiently detected by FASER's front veto, while neutrino interactions are relatively rare and typically have different kinematics.  The FASER Collaboration presented their first analysis on dark photons, providing a two-track signal, in which they accounted for a variety of possible background sources associated with veto inefficiencies, neutral hadrons, muons missing the veto, neutrinos, and non-collision background~\cite{FASER:2023tle}. These backgrounds were determined to be either very small or negligible, and no events with two reconstructed tracks passing the veto requirement have been observed. We therefore assume that backgrounds for multi-track signatures can be considered negligible. 

More recently, the FASER Collaboration also presented results for a search for axion-like particles, considering a multi-photon signal~\cite{Kock:2892328}. A potentially sizable background from neutrinos interacting at the end of the detector was identified, particularly at energies below 1~TeV. To address this issue, a high-precision preshower upgrade is planned to be installed at the end of 2024, which will allow the experiment to identify multi-photon signatures and differentiate them from neutrino backgrounds~\cite{Boyd:2803084}. We therefore assume that backgrounds for multi-photon signatures are also negligible. 

The design for FASER2 is currently under development~\cite{Feng:2022inv}. It will have a similar conceptual architecture as the currently-operating FASER detector, and background rejection is one of the key considerations for the detector design. Although the design has not yet been finalized, it is envisioned that the same searches will be background free.

\section{Reach} \label{sec:reach}

\begin{figure*}[!t]
\centering
\includegraphics[width=.8\textwidth]{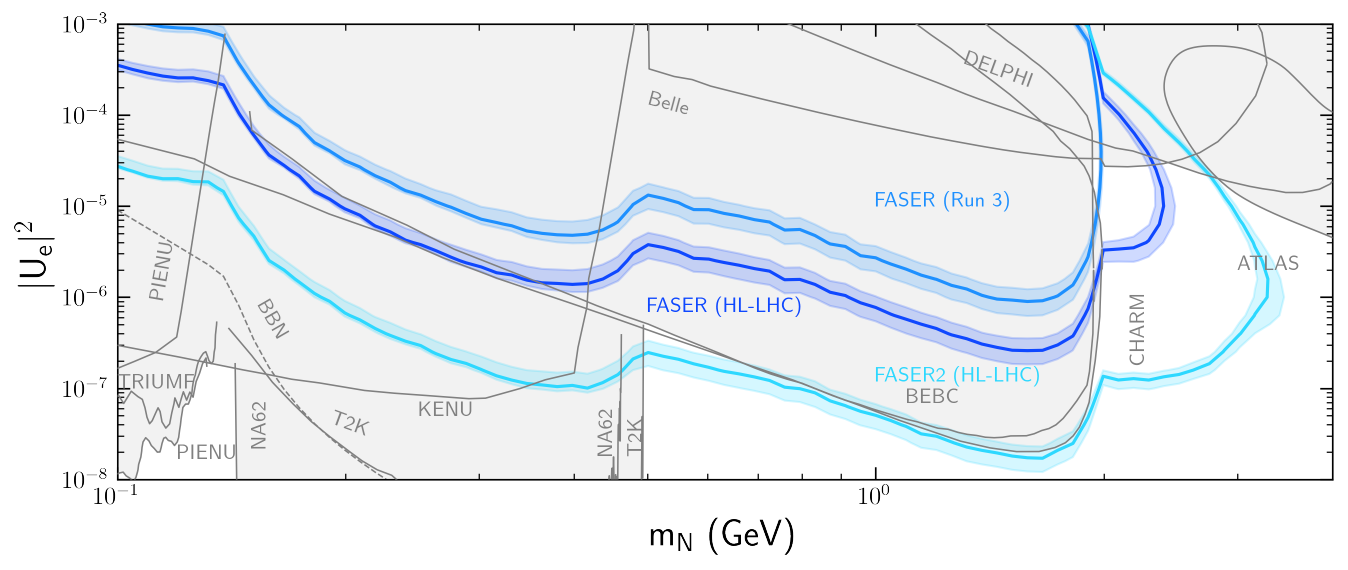}
\includegraphics[width=.8\textwidth]
{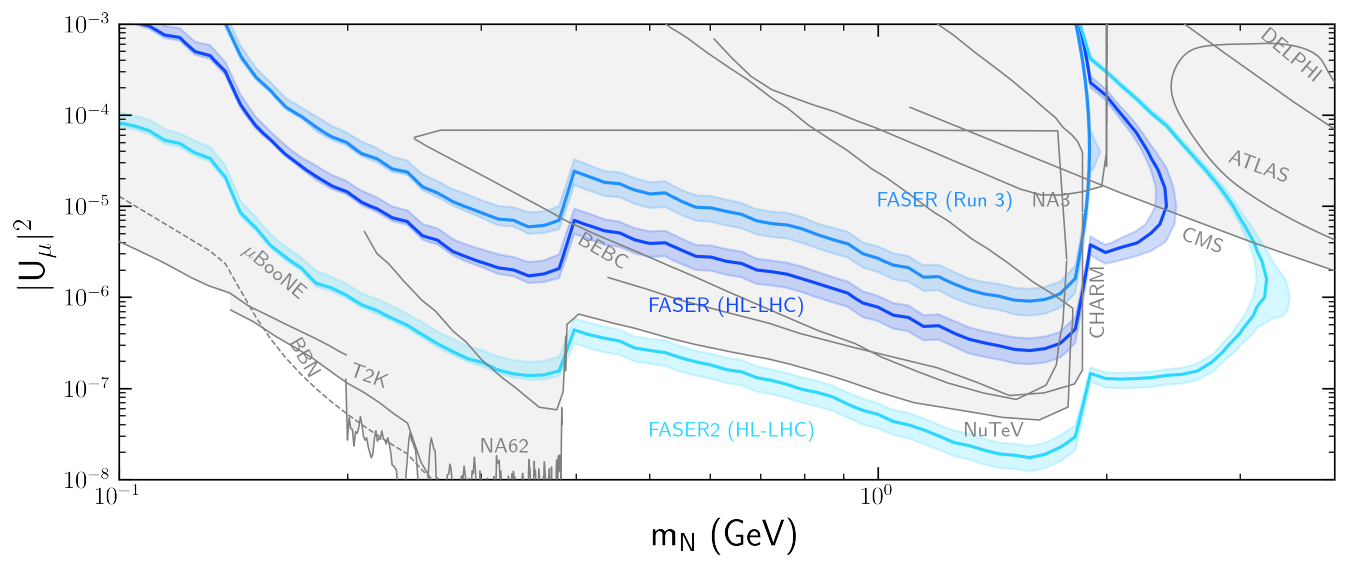}
\includegraphics[width=.8\textwidth]{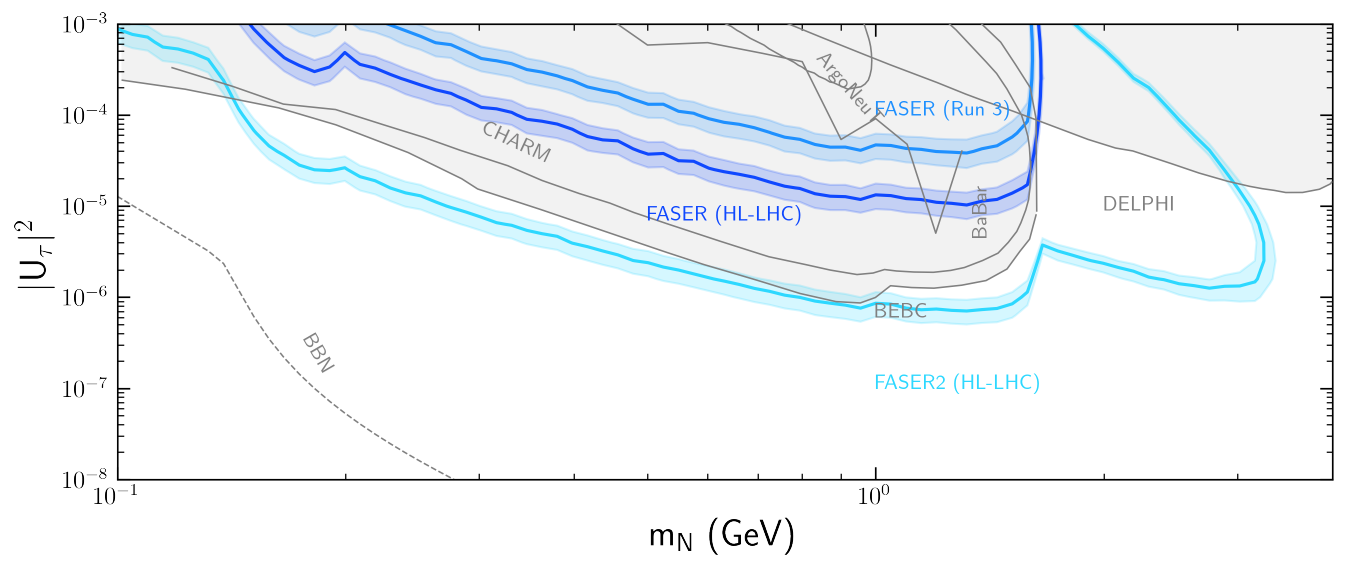}
\caption{$N=3$ signal event contours for the single-coupling 100 (top), 010 (middle), and 001 (bottom) benchmark models. Each blue contour corresponds to a detector/collider configuration given in \cref{tab:geom}. Current bounds from other particle experiments exclude the gray shaded regions, and BBN constraints exclude the region below the dashed contour. All of the production modes in   \cref{tab:production modes} are included, and all of the decay modes in \cref{tab:decay modes} are included, except for the invisible $\nu \bar{\nu} \nu$ modes. For the FASER and FASER2 contours, a momentum cut of $p_N > 100~\gev$ has been applied.}
    \label{fig:sensitivity simple benchmarks}
\end{figure*}

\begin{figure*}[!t]
    \centering
    \includegraphics[width=.8\textwidth]{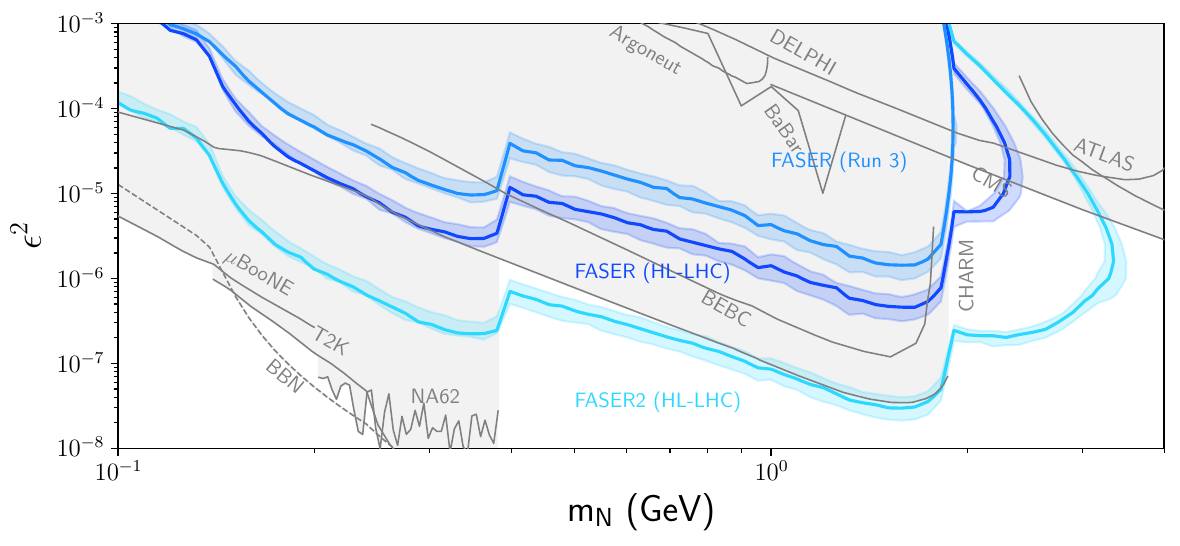}
    \includegraphics[width=.8\textwidth]{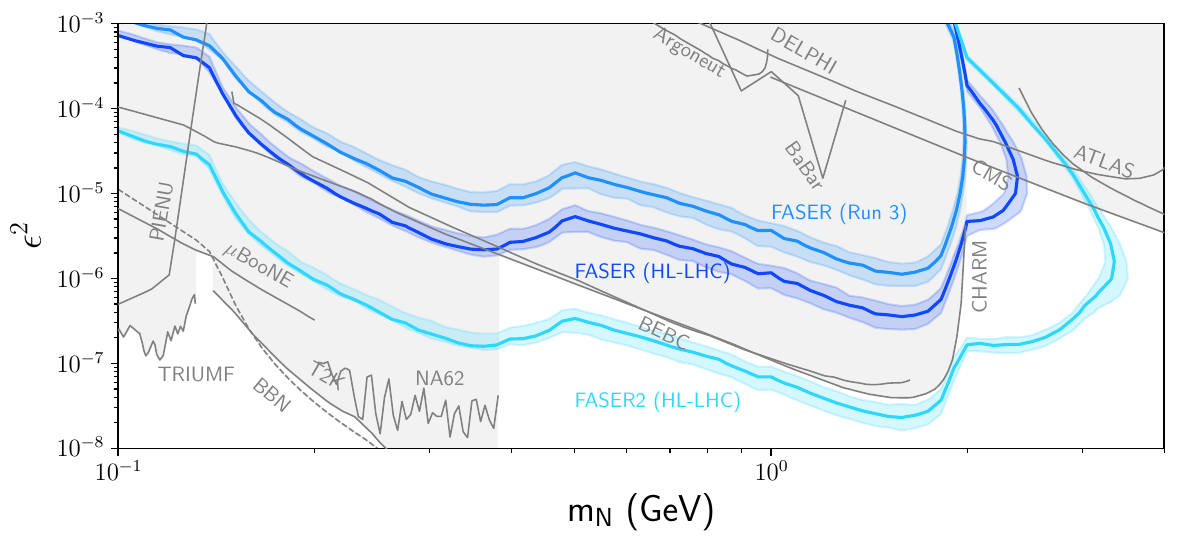}
\caption{As in \cref{fig:sensitivity simple benchmarks}, but for the mixed-coupling 011 (top) and 111 (bottom) benchmark models.  In contrast to the single-coupling models, constraints from other experiments are not available in the literature.  The current bounds shown in the gray region are roughly estimated based on a method described in the text.}
    \label{fig:sensitivity new benchmarks}
\end{figure*}

We now present the $N=3$ signal event contours in the $(m_N, \epsilon)$ plane for the various benchmark scenarios.  In the absence of background, these may be considered to be discovery contours. The results for the 100, 010, and 011 benchmark models are given in \cref{fig:sensitivity simple benchmarks}, and the results for the 011 and 111 benchmark models are given in \cref{fig:sensitivity new benchmarks}.  The signal event rates include all of the production modes listed in \cref{tab:production modes} and all of the decay modes shown in \cref{tab:decay modes}, except for the invisible decay modes $\nu \bar{\nu} \nu$. Additionally, a momentum cutoff of $p_N>100$ GeV is applied.  The event rates are determined by setting $\texttt{nsample} = 25$ in the \texttt{FORESEE} MC simulation~\cite{Kling:2021fwx}, and the $(m_N, \epsilon)$ parameter space is scanned using 100 equally log-spaced masses $m_N$ from 100 MeV to 4 GeV, and 50 equally log-spaced couplings $\epsilon$ from $10^{-5}$ to $1$. 

Results are presented for the three FASER/FASER2 configurations shown in \cref{tab:geom}. The shaded band corresponds to the flux uncertainty, following the prescription in \cref{subsec: HNL production Rates}. Sensitivity contours for the 100, 010, and 001 models have been investigated previously~\cite{Kling:2018wct,FASER:2018eoc}. 
Our obtained sensitivities are consistent with these results over most of the mass range. They are slightly improved for $m_N \alt m_K$, due to an issue that was found in the kaon decay modes of these previous results.  Sensitivity reaches for FASER and FASER2 for the 011 and 111 models have not been determined before.

Various threshold effects can be observed in the sensitivity contours, depending on which production channels dominate HNL production. In the 100 benchmark, the production of HNLs is predominantly governed by pions in the region $100~\mev \leq \ m_N \leq \ m_{\pi}$, transitions to being dominated by kaons in the range $m_{\pi} \leq \ m_N \leq \ m_K$, shifts to being influenced by $D$ and $\tau$ particles within the interval $m_K \leq \ m_N \leq \ m_D$, and eventually becomes dominated by $B$ mesons for $m_N \ \geq \ m_D$. Additionally, one can infer from the sensitivity contours that the 011 benchmark sensitivity is primarily driven by the $U_\mu$ coupling and the 111 benchmark sensitivity is primarily driven by the $U_e$ coupling.

It is, of course, interesting to compare the sensitivity of FASER and FASER2 to current constraints.  For the 100, 010, and 001 models, the exclusion bounds have been determined for many past experiments, and are available using the Python package \texttt{HNL-Limits}~\cite{Fernandez-Martinez:2023phj}. For the 100 and 001 benchmarks, the BEBC and CHARM bounds in \texttt{HNL-Limits} have been refined in Ref.~\cite{Barouki:2022bkt}, and we have adopted those limits. For the 010 benchmark, we use the lower bound provided by the CHARM Collaboration~\cite{CHARM:1985nku}. To establish the upper limit for this benchmark, we use the bounds derived for the 100 benchmark model from the CHARM recasting detailed in Ref.~\cite{Barouki:2022bkt} as an approximation. We further apply a cutoff at $m_{D_s} - m_{\mu}$ in the 010 and 011 CHARM bounds to account for the allowed phase space.  In addition, for small masses, there are constraints from Big Bang nucleosynthesis (BBN), which constrains the lifetime of the HNL to be less than 0.1~s~\cite{Dolgov:2000jw}.

For the 011 and 111 models, the sensitivities of previous experiments are not often found in the literature.  To compare FASER and FASER2 to existing constraints, an analysis similar to the one we have done for FASER and FASER2 must be done, requiring a dedicated analysis for each model and experiment.  This is beyond the scope of this work.  However, we may roughly estimate the sensitivities of previous experiments in the 011 and 111 models by extrapolating their limits in the 100, 010, and 001 models, as we now describe.

In peak search experiments, one looks for evidence of HNL production by searching for peaks in the energy spectrum of co-produced particles. One such example are the bounds placed by the PIENU Collaboration~\cite{PIENU:2017wbj}, where they performed a search for peaks in the positron energy spectrum in the process $\pi^+\to N e^+$. Bounds in these types of experiments place limits on $U_\alpha$ that are independent of the HNL lifetime and decay products and thus will not be effected by the introduction of mixed couplings. The bounds on $U_\alpha$ can therefore be re-scaled directly to obtain bounds for $\epsilon$. 

In prompt decay or decay-in-flight searches, such as those at ATLAS~\cite{ATLAS:2022atq} or CHARM~\cite{Boiarska:2021yho}, one searches for evidence of HNLs by looking for HNL decay products within the detector.  In these experiments, the introduction of mixed couplings can effect the kinematic distributions of final states, as well as the HNL lifetime, causing a change in the overall expected event rate. As a very rough approximation, we can neglect the effect on the kinematics by making the simplifying assumptions that the detector acceptance and efficiency are uniform across all benchmarks. In the long lifetime limit, then, the expected event rate in these experiments scales as 
\begin{equation}
    N_{\alpha} \propto \frac{|U_{\alpha}|^2}{\tau} \ ,
\end{equation}
where $N_\alpha$ is the contribution to the total event rate occurring through $\nu_\alpha$ mixing. This allows one to relate the expected event rates between the mixed coupling and single coupling scenario by
\begin{equation}
    \frac{N_{\alpha,m}}{N_{\alpha,s}} = \frac{|U_{\alpha,m}|^2}{|U_{\alpha,s}|^2}\frac{\tau_s}{\tau_m} \ ,
\end{equation}
where the subscript $m$ ($s$) denotes the mixed (single) coupling scenario. Approximate exclusion bounds derived in this way for the 011 and 111 benchmarks are plotted in \cref{fig:sensitivity new benchmarks}.

For the sensitivity reaches in the single coupling scenarios shown in \cref{fig:sensitivity simple benchmarks}, the results may be summarized as follows.  For FASER (Run 3), the reach beyond current bounds is rather modest and is limited to the 100 model.  For FASER (HL-LHC), the reach is extended, with discovery prospects in the 100 and 010 scenarios for $m_N \sim 2~\gev$.  Finally, for FASER2 (HL-LHC), there are significant regions of parameter space in all 3 models with sensitivity beyond current bounds. This includes regions of parameter space with $10^{-7} \alt \epsilon^2 \alt 10^{-5}$ and $2~\gev \alt m_N \alt 4~\gev$, where the HNLs are produced in $B$ meson decays, and also regions of low $\epsilon$ at lower $m_N$.  This is especially notable in the 001 scenario, where the new parameter space probed extends all the way down to HNL masses of approximately 150 MeV.

For the sensitivity contours in mixed coupling scenarios shown in \cref{fig:sensitivity new benchmarks}, we find that the comparison of the reach of FASER and FASER2 relative to current bounds for the 011 and 111 scenarios approximately mirrors those for the 010 and 100 models, respectively.  In particular, for FASER2 (HL-LHC) there is again new parameter space probed with $10^{-7} \alt \epsilon^2 \alt 10^{-5}$ and $2~\gev \alt m_N \alt 4~\gev$, and also improved sensitivity over current bounds at low $\epsilon$ and masses $400~\mev \alt m_N \alt 2~\gev$. We reiterate, however, that in this comparison, the FASER and FASER2 bounds derived here are compared to current bounds that have been derived in the very rough way we have outlined above.  Detailed comparisons require dedicated analyses of the reach of other experiments in these less minimal, mixed-coupling scenarios.

Finally, let us comment on the flux uncertainties, which are shown as shaded bands in \cref{fig:sensitivity simple benchmarks,fig:sensitivity new benchmarks}. Despite substantial flux uncertainties, which can be as large as a factor of two for charm production, their overall impact on the sensitivity is relatively small. This is due to a strong coupling dependence of the event rate at both small and large couplings. The flux uncertainties mainly affect the reach at the higher mass edge of the sensitivity contour. We note that further measurement, especially the measurement of collider neutrino fluxes, will help to further decrease those flux uncertainties.

\section{Conclusions} \label{sec:conclusions}

HNLs are well motivated in extensions of the SM designed to address the outstanding puzzles of neutrino masses, dark matter, and baryogengesis. In this work, we have comprehensively studied the possibility of probing HNLs with mass in the 100 MeV to 10 GeV range and completely arbitrary mixings with the three active SM neutrinos.  We have produced a comprehensive package, \texttt{HNLCalc}, that computes the properties of the HNL model, including hundreds of production and decay modes. Using \texttt{HNLCalc}, we created an HNL module in the \texttt{FORESEE} simulation package to evaluate the discovery potential of HNLs in various experimental setups.

In particular, we have estimated the sensitivity to HNLs for FASER in Run 3, FASER at the HL-LHC, and FASER2 at the HL-LHC. As an illustration of the flexibility of \texttt{HNLCalc}, we have  considered five benchmark models.  For the well-studied 100, 010, and 001 benchmarks, we find that FASER in Run 3 has rather limited discovery prospects.  However, FASER at the HL-LHC can probe new parameter space with $m_N \sim 2~\gev$, and FASER2 at the HL-LHC can probe new parameter space for a wide range of HNL masses from 150 MeV to 4 GeV.

Additionally, we have considered two new mixed-coupling benchmark models, 011 and 111, and determined the sensitivity of FASER and FASER2.  These new benchmarks are not well studied and there are no comprehensive studies of the sensitivity of other current and proposed experiments in these models.  We have roughly estimated the sensitivity of past experiments, and we find that FASER2 at the HL-LHC can likely probe new parameter space for a wide range of HNL masses from 400 MeV to 4 GeV.

Overall, this study contributes significantly to the understanding of HNLs and their potential implications in particle physics. We have produced \texttt{HNLCalc}, a flexible, fast, comprehensive, and publicly-available tool for computing HNL decay and production rates with arbitrary couplings. In this study, we have used this tool to extend the \texttt{FORESEE} simulation package to incorporate HNLs.  Models with general HNL couplings are more complicated than models with a single coupling, but they are also more well-motivated, and studies of the discovery prospects of other experiments in these models are encouraged.

\section*{Acknowledgements}

We are grateful to the authors and maintainers of many open-source software packages, including scikit-hep~\cite{Rodrigues:2020syo} and CRMC~\cite{crmc201}. This work is supported in part by U.S.~National Science Foundation (NSF) Grant PHY-2210283 and Simons Foundation Grant 623683.  The work of J.L.F.~is supported in part by NSF Grant PHY-2111427, Simons Investigator Award \#376204, and Heising-Simons Foundation Grants 2019-1179 and 2020-1840.  F.K.~acknowledges support by the Deutsche Forschungsgemeinschaft under Germany’s
Excellence Strategy---EXC 2121 Quantum Universe---390833306.

\onecolumngrid

\appendix

\section{HNL Production Modes}
\label{apx:production}

In this appendix, we present a comprehensive compilation of expressions for the leading processes for HNL production. Specifically, in the following five subsections we present the branching fractions for the processes $P \to l N$, $P \to P^\prime l N$, $P \to V l N$, $\tau \to PN$, and $\tau \to \nu l N, \bar{\nu} l N$, where $P$ and $P^\prime$ denote pseudoscalar mesons, $V$ denotes vector mesons, $l$ is a charged lepton, and $N$ is the HNL. The decays implemented are shown in \cref{tab:production modes}.

\subsection{2-Body Pseudoscalar Decays \boldmath{$P \to l N$}}
\label{sbsec:2bdy_pseudo}

The 2-body leptonic decays of pseudoscalar mesons into HNLs have branching fractions~\cite{Shrock:1980vy,Gorbunov:2007ak}
\begin{equation}
\begin{split}
B(P \to l_{\alpha} N ) = & \ \tau_P \ |U_{\alpha}|^2 \ \frac{G_F^2 m_P m_N^2 |V_P|^2 f_P^2}{8 \pi}  \left[ 1- \frac{m_N^2}{m_P^2} + 2 \frac{m_{l}^2}{m_P^2}+ \frac{m_{l}^2}{m_N^2} \left( 1- \frac{m_{l}^2}{m_P^2} \right) \right] \\
& \times \sqrt{\left(1+ \frac{m_N^2}{m_P^2}- \frac{m_{l}^2}{m_P^2}\right)^2 - 4 \frac{m_N^2}{m_P^2}} \ ,
\end{split}
\end{equation}
where $\alpha = \ e, \mu, \tau$; $\tau_P$ is the meson lifetime; $U_{\alpha}$ parameterizes the mixing with the active neutrino; and $G_F$ is the Fermi constant. The decay constants $f_P$ are given in \cref{tab:decay constants}, and $V_P$ denotes the relevant CKM matrix element; for example, for $P= \ \pi^+$, $V_P= \ V_{ud}$. 

\begin{table}[bp]
    \centering
    \begin{tabular}{|l|D{.}{.}{3}|l|D{.}{.}{2}|}
        \hline
        \quad $P$                       & \multicolumn{1}{c|}{$f_P \ (\text{MeV})$} & \quad $V$   & \multicolumn{1}{c|}{$f_V \ (\text{MeV})$} \\
        \hline
        $\pi^0$~\cite{Chang:2018aut}    &  130.3             & $\rho^0$~\cite{Ebert:2006hj}   &  220 \\
        \hline
        $\pi^+$~\cite{Chang:2018aut}    &  130.3              & $\rho^+$~\cite{Ebert:2006hj}  & 220 \\ 
        \hline
        $K^+$~\cite{Chang:2018aut}      & 156.4              &  $\omega$~\cite{Ebert:2006hj}  & \quad 195 \\
        \hline
        $\eta$~\cite{Feldmann:1999uf}  & 78.4              &  ${K^*}^+$~\cite{Chang:2018aut} & 204 \\      
        \hline
        $\eta'$~\cite{Feldmann:1999uf}  & -95.7                & $\phi$~\cite{Chang:2018aut}                               & 229         \\
        \hline
        $D^+$~\cite{CLEO:2005jsh}       &   222.6              &                                &\\
        \hline
        $D_s^+$~\cite{Stone:2006cc}     & 280.1              &                                &  \\
        \hline
        $B^+$~\cite{Gorbunov:2007ak}    & 190                &                                & \\
        \hline
        $B_c^+$~\cite{Gorbunov:2007ak}  & 480                &                                & \\
        \hline
    \end{tabular}
    \caption{The pseudoscalar and vector meson decay constants used in this study.}
    \label{tab:decay constants}
\end{table}

\subsection{3-Body Pseudoscalar Decays \boldmath{$P \to P' l N$}}
\label{sbsec:3bdy_pseudo}

For pseudoscalar mesons decaying semi-leptonically to pseudoscalar mesons, the differential branching fraction is~\cite{Gorbunov:2007ak}
\begin{equation}
\begin{split}
\frac{d B\left(P \to P^{\prime} l_\alpha N\right)}{d E_N d q^2} = & \ \tau_P \ \left|U_\alpha\right|^2 \ \frac{G_F^2 \left|V_{P P^{\prime}}\right|^2}{64 \pi^3 m_P^2} \ c_P \  \left\{ f_{-}^2\left(q^2\right) \cdot\left[q^2\left(m_N^2+m_l^2\right)-\left(m_N^2-m_l^2\right)^2\right] \right. \\
&+2 f_{+}\left(q^2\right) f_{-}\left(q^2\right)\left[m_N^2\left(2 m_P^2-2 m_{P^{\prime}}^2-4 E_N m_P-m_l^2+m_N^2+q^2\right) \right.\\
&\left. +m_l^2\left(4 E_N m_P+m_l^2-m_N^2-q^2\right)\right] \\
&+f_{+}^2\left(q^2\right)\left[\left(4 E_N m_P+m_l^2-m_N^2-q^2\right)\left(2 m_P^2-2 m_{P^{\prime}}^2-4 E_N m_P-m_l^2+m_N^2+q^2\right) \right. \\
&\left. \left. -\left(2 m_P^2+2 m_{P^{\prime}}^2-q^2\right)\left(q^2-m_N^2-m_l^2\right) \right] \right\} \ ,
\label{eq:PtoP'}
\end{split}
\end{equation}
where $q^2 = (p_{l} + p_N)^2$, $E_N$ is the HNL energy in the $P$ COM frame, and $V_{P P'}$ is the appropriate CKM matrix element for the process (e.g., for $D^+ \to \overline{K}^0 l^+ N$, $V_{P P'} = \  V_{cs}$). The constant $c_P$ and the form factors $f_-(q^2)$ and $f_+(q^2)$ are defined below. \cref{eq:PtoP'} corrects a minor typo in Eq.~(B2) of Ref.~\cite{Gorbunov:2007ak}, which was missing a plus sign. Additionally, we replace the masses $m_K$ and $m_{\pi}$ in Eq.~(B2) of Ref.~\cite{Gorbunov:2007ak} with $m_P$ and $m_{P'}$, respectively. 

To determine the branching fractions we must integrate \cref{eq:PtoP'} over the region~\cite{ParticleDataGroup:2018ovx}
\begin{align}
\begin{split}
    (m_{l} + m_N)^2 & \leq \ q^2 \  \leq \ (m_H - m_{H'})^2 \ , \\
     E_N(m_{N H'}^{2\, \text{min}},q^2) & \leq \ E_N \leq \ E_N(m_{N H'}^{2\, \text{max}},q^2) \ ,
\label{Ebounds}
\end{split}
\end{align}
where
\begin{align}
\begin{split}
E_N (m_{N H'}^2, q^2) = & \ \frac{q^2 + m_{N H'}^2 - m_{H'}^2 - m_{l}^2}{2 m_H} \ , \\
m_{N{H'}}^{2\, \text{min}} = \ & \left(E_N^* + E_{H'}^* \right)^2 - \left( \sqrt{E_N^{*2} - m_N^2} + \sqrt{E_{H'}^{*2} - m_{H'}^2} \right)^2 \ ,\\
m_{N{H'}}^{2\, \text{max}} = & \  \left(E_N^* + E_{H'}^* \right)^2 - \left(\sqrt{E_N^{*2} - m_N^2} - \sqrt{E_{H'}^{*2} - m_{H'}^2} \right)^2 \ , \\
\end{split}
\end{align}
where $E_N^* = ( m_{lN}^2 - m_l^2 + m_N^2)/2 m_{lN}$, and $E_{H'}^* = (m_H^2 - m_{lN}^2 - m_{H'}^2)/2 m_{lN}$ are the energies of $N$ and $H'$ in the COM frame of $l-N$ system.  \cref{Ebounds} corrects a typo found below Eq.~(2) of Ref.~\cite{Graverini:2133817}, where the $q^2$ dependence of the integration bounds for $E_N$ was neglected.

The hadronic form factors  $f_-(q^2)$ and $f_+(q^2)$ appearing in \cref{eq:PtoP'} are defined by~\cite{Graverini:2133817}
\begin{align}
f_{+}\left(q^2\right) =  & \ \frac{f_{+}(0)}{\left(1-q^2 / m_{V'}^2\right)}\ , \label{eq:pseudo form factors1} \\
f_0\left(q^2\right)  = & \ \frac{f_0(0)}{\left(1-q^2 / m_S^2\right)} \ , \label{eq:pseudo form factors2} \\
f_0\left(q^2\right)  = & \ f_{+}\left(q^2\right) +\frac{q^2}{m_P^2-m_{P'}^2} f_{-}\left(q^2\right) \ . \label{eq:pseudo form factors3}
\end{align}
Here $m_{V^\prime}$ and $m_S$ are the masses of the vector and scalar resonances, respectively. These are determined from the quark transition of the decay (e.g., for $D^+ \to \overline{K}^0 l^+ N$, which involves a $c \to s$ quark transition, $m_{V^\prime}  = \ m_{D_s^{*+}}$, and $m_S  =  \ m_{D_s^+}$). It is important to note that $f_+(0)$ and $f_-(0)$ are determined by \cref{eq:pseudo form factors1,eq:pseudo form factors2,eq:pseudo form factors3}, and therefore, we only provide values for $f_0(0)$, which are listed in~\cref{tab:pseudo form factors}.

The constants $c_P$ are determined by the quark content of the initial and final state mesons. As an example, consider the decay $D^+ \to \eta l^+ N$, which depends on the matrix element $\langle \eta | \bar{d} \gamma^{\mu} P_L c | D^+ \rangle$. The quark content of $\eta$ and $\eta^\prime$ is related to $\eta_8$ and $\eta_1$ through a rotation matrix:
\begin{equation}
\begin{pmatrix}
\eta \\ \eta^\prime \end{pmatrix} = \ \begin{pmatrix} \cos \theta_P & - \sin \theta_P \\ \sin \theta_P & \cos \theta_P \end{pmatrix} \begin{pmatrix} \eta_8 \\ \eta_1 
\end{pmatrix} \ .
\end{equation}
Inserting $\eta_8 = \ ( u \bar{u} + d \bar{d} - 2 s \bar{s})/\sqrt{6} $ and $\eta_1 = \  ( u \bar{u} + d \bar{d} + s \bar{s})/\sqrt{3}$ we obtain,
\begin{equation}
    \eta = \ \left( \frac{\cos \theta_P}{\sqrt{6}} - \frac{\sin \theta_P}{\sqrt{3}} \right) u \bar{u} + \left( \frac{\cos \theta_P}{ \sqrt{6}} - \frac{\sin \theta_P}{\sqrt{3}} \right) d \bar{d}  +\left(  - \frac{2 \cos \theta_P}{\sqrt{6}} - \frac{\sin \theta_P}{ \sqrt{3}} \right) s \bar{s} \ .
\end{equation}
From this, we observe that the matrix element $\langle \eta | \bar{d} \gamma^{\mu} P_L c | D^+ \rangle$ is proportional to $\left( \frac{\cos \theta_P}{\sqrt{6}} - \frac{\sin \theta_P}{\sqrt{3}} \right) \langle d | \bar{d} \gamma^{\mu} P_L c | c \rangle$, where $\langle d | \bar{d} \gamma^{\mu} P_L c | c \rangle$ is the same matrix element encountered in the decay $D^0 \to \pi^-  l^+  N$ and thus shares the same form factors. This implies that the parameter $c_P$ for the decay $D^+ \to \eta  l^+ N$ is $\left( \frac{\cos \theta_P}{\sqrt{6}} - \frac{\sin \theta_P}{\sqrt{3}} \right)^2$. 
The values of $c_P$ are listed in~\cref{tab:pseudo form factors}.

\begin{table}[bp]
    \centering
    \begin{tabular}{|l|D{.}{.}{4}|c|}
        \hline
        Decay Channel                & \multicolumn{1}{c|}{$f_0(0)$} & $c_P$\\
        \hline
        $K^+ \to \pi^0$~\cite{Carrasco:2016kpy}            & 0.970    & 1/2 \\
        \hline
        $K_S \to \pi^+$~\cite{Aoki:2017spo}    & 0.9636    & 1/2 \\
        \hline
        $K_L \to \pi^+$ \cite{Aoki:2017spo}    & 0.9636    & 1/2 \\
        \hline
        $\overline{D}^0 \to \pi^+$     \cite{Melikhov:2000yu}         & 0.69     & 1  \\
        \hline
        $\overline{D}^0 \to K^+$ \cite{Carrasco:2015bhi} & 0.747    & 1\\
        \hline
        $D^+ \to \pi^0$    \cite{Melikhov:2000yu}          & 0.69    & 1/2 \\
        \hline
        $D^+ \to \eta$     \cite{Melikhov:2000yu}          & 0.69    & $( \frac{\cos \theta_P}{\sqrt{6}} - \frac{\sin \theta_P}{\sqrt{3}})^2$ \\
        \hline
        $D^+ \to \eta^\prime$    \cite{Melikhov:2000yu}          &  0.69    & $( \frac{\sin \theta_P}{\sqrt{6}} + \frac{\cos \theta_P}{\sqrt{3}})^2$ \\
        \hline
        $D^+ \to \overline{K}^0$ \cite{Carrasco:2015bhi} & 0.747    & 1\\     
        \hline
        $D_s^+ \to \overline{K}^0$ \cite{Carrasco:2015bhi} & 0.747    & 1\\
        \hline
        $D_s^+ \to \eta$   \cite{Duplancic:2015zna}          & 0.495    & 1   \\
        \hline
        $D_s^+ \to \eta^\prime$    \cite{Duplancic:2015zna}        & 0.557    & 1   \\
        \hline
        $B^+ \to \pi^0$     \cite{Melikhov:2000yu}         & 0.29    & 1/2  \\
        \hline
        $B^+ \to \eta$    \cite{Melikhov:2000yu}           & 0.29    & $( \frac{\cos \theta_P}{\sqrt{6}} - \frac{\sin \theta_P}{\sqrt{3}})^2$  \\
        \hline
        $B^+ \to \eta^\prime$   \cite{Melikhov:2000yu}           & 0.29    & $( \frac{\sin \theta_P}{\sqrt{6}} + \frac{\cos \theta_P}{\sqrt{3}})^2$  \\
        \hline
        $B^+ \to \overline{D}^0$ \cite{Na:2015kha}      & 0.66     & 1 \\ 
        \hline
        $B^0 \to \pi^+$   \cite{Melikhov:2000yu}           & 0.29    & 1  \\
        \hline
        $B^0 \to D^-$ \cite{Na:2015kha}      & 0.66     & 1 \\ 
        \hline
        $B_s^0 \to K^-$    \cite{Melikhov:2000yu}          & 0.31    &  1  \\
        \hline
        $B^0_s \to D_s^-$ \cite{Chen:2011ut}            & -0.65    & 1 \\    
        \hline
        $B_c^+ \to D^0$ \cite{Ivanov:2000aj}             & 0.69    & 1 \\     
        \hline
        $B_c^+ \to \eta_c$ \cite{Ivanov:2000aj}          & 0.76    & 1   \\
        \hline
        $B_c^+ \to B^0$ \cite{Ivanov:2000aj}           & -0.58    & 1\\     
        \hline
        $B_c^+ \to B_s^0$  \cite{Ivanov:2000aj}           & -0.61    & 1 \\    
        \hline  
    \end{tabular}
\caption{The parameters $f_0(0)$ and $c_P$, which appear in the $P \to P^\prime l N$ branching fraction expressions of \cref{eq:PtoP',eq:pseudo form factors1,eq:pseudo form factors2,eq:pseudo form factors3}, where $\theta_P= -11.5^{\circ}$~\cite{ParticleDataGroup:2020ssz}.}
    \label{tab:pseudo form factors}
\end{table}

\subsection{3-Body Pseudoscalar Decays \boldmath{$P \to V l N$}}
\label{sbsec:3bdy_vector}

For pseudoscalar mesons decaying semi-leptonically to vector mesons, the differential branching fraction is~\cite{Gorbunov:2007ak}
\begin{equation}
\begin{split}
\frac{d B\left(P \to V l_\alpha N\right)}{d E_N d q^2}  = & \ \tau_P \ \left|U_\alpha\right|^2 \frac{G_F^2 \left|V_{P V}\right|^2}{32 \pi^3 m_P} \ c_V \ \left\{\frac{f_2^2}{2}\left(q^2-m_N^2-m_l^2+\omega^2 \frac{\Omega^2-\omega^2}{m_V^2}\right)\right. \\
&+\frac{f_5^2}{2}\left(m_N^2+m_l^2\right)\left(q^2-m_N^2+m_l^2\right)\left(\frac{\Omega^4}{4 m_V^2}-q^2\right)\\
&+2 f_3^2 m_V^2\left(\frac{\Omega^4}{4 m_V^2}-q^2\right)\left(m_N^2+m_l^2-q^2+\omega^2 \frac{\Omega^2-\omega^2}{m_V^2}\right)\\
&+2 f_3 f_5\left[m_N^2 \omega^2+\left(\Omega^2-\omega^2\right) m_l^2\right]\left(\frac{\Omega^4}{4 m_V^2}-q^2\right)+2 f_1 f_2\left[q^2\left(2 \omega^2-\Omega^2\right)+\Omega^2\left(m_N^2-m_l^2\right)\right]\\ 
&+\left. f_1^2\left[\Omega^4\left(q^2-m_N^2+m_l^2\right)-2 m_V^2\left[q^4-\left(m_N^2-m_l^2\right)^2\right]+2 \omega^2 \Omega^2\left(m_N^2-q^2-m_l^2\right)+2 \omega^4 q^2\right]\right\}\\
&+\frac{f_2 f_5}{2}\left[\omega^2 \frac{\Omega^2}{m_V^2}\left(m_N^2-m_l^2\right)+\frac{\Omega^4}{m_V^2} m_l^2+2\left(m_N^2-m_l^2\right)^2-2 q^2\left(m_N^2+m_l^2\right)\right]\\
&+f_2 f_3\left[\Omega^2 \omega^2 \frac{\Omega^2-\omega^2}{m_V^2}+2 \omega^2\left(m_l^2-m_N^2\right)+\Omega^2\left(m_N^2-m_l^2-q^2\right)\right] \ ,
\end{split}
\label{eq:3bodypseudoscalar}
\end{equation}
where $\omega^2 = m_P^2-m_V^2+m_N^2-m_l^2-2 m_P E_N$, and $\Omega^2= \ m_P^2-m_V^2-q^2$. The constants $c_V$ are conceptually identical to the constants $c_P$ discussed in \cref{sbsec:3bdy_pseudo}.

The form factors are
\begin{align}
f_1 = & \ \frac{V}{m_P+m_V} \ , \\
f_2 = & \ \left(m_P+m_V\right) A_1 \ , \\
f_3 = & \  -\frac{A_2}{m_P+m_V} \ , \\
f_4 = & \ \left[ m_V\left(2 A_0-A_1-A_2\right)+m_P\left(A_2-A_1\right)\right] \frac{1}{q^2} \ , \\
f_5 = & \ f_3+f_4 \ .
\end{align}
$A_0$ and $V$ have the form
\begin{equation}
f\left(q^2\right)= \ \frac{f(0)}{\left(1-q^2 / m^2\right)\left(1-\sigma_1 q^2 / m^2+\sigma_2 q^4 / m^4\right)} \ ,
\label{eq:vec form factors1}
\end{equation}
where $m= \ m_S$ for $A_0$, and $m= \ m_{V^\prime}$ for $V$. $A_1$ and $A_2$ have the form
\begin{equation}
f\left(q^2\right) = \ \frac{f(0)}{\left(1-\sigma_1 q^2 / m_{V^\prime}^2+\sigma_2 q^4 / m_{V^\prime}^4\right)} \ .
\label{eq:vec form factors2}
\end{equation}

The parameters $m_S$ and $m_{V^\prime}$ are determined from the decay mode in a manner identical to $P \to P^\prime l N$.  The parameters $\sigma_1$ and $\sigma_2$ are unique for each of the form factors $A_0$, $A_1$, $A_2$, and $V$. A summary of the form factor parameters can be found in \cref{tab:vector form factors}.

The form factors for certain $B_c$ decays are a special case and all of the form factors $A_0$, $A_1$, $A_2$, and $V$ are chosen to have the form~\cite{Ivanov:2000aj}

\begin{equation}
    f \left( q^2 \right)  = \ \frac{f(0)}{1-q^2/m_{\text{fit}}^2-\delta(q^2/m_{\text{fit}}^2)^2} \ ,
    \label{eq:B_c form factors}
\end{equation}
where $\delta$ and $m_{\text{fit}}$ are fitting parameters unique to each of the form factors $A_0$, $A_1$, $A_2$, and $V$. These fitting parameters for the relevant decays can be found in \cref{tab:vector form factors B_c}.
\begin{table}[bp]
    \centering
    \begin{tabular}{|l|S|S|S|S|S|S|S|S|S|S|S|S|c|}
    \hline
                         & \multicolumn{3}{c|}{$A_0$}        & \multicolumn{3}{c|}{$A_1$}          & \multicolumn{3}{c|}{$A_2$}          & \multicolumn{3}{c|}{$V$}         &    \\
        \hline
        Decay Channel    & \multicolumn{1}{c|}{$f_0(0)$}  & \multicolumn{1}{c|}{$\sigma_1$} & \multicolumn{1}{c|}{$\sigma_2$} & \multicolumn{1}{c|}{$f_0(0)$} & \multicolumn{1}{c|}{$\sigma_1$} & \multicolumn{1}{c|}{$\sigma_2$} & \multicolumn{1}{c|}{$f_0(0)$} & \multicolumn{1}{c|}{$\sigma_1$} & \multicolumn{1}{c|}{$\sigma_2$} & \multicolumn{1}{c|}{$f(0)$} & \multicolumn{1}{c|}{$\sigma_1$} & \multicolumn{1}{c|}{$\sigma_2$} & $c_V$\\
        \hline
        $D^0 \to \rho^-$ \cite{Melikhov:2000yu} 
                         & 0.66     & 0.36       & 0          & 0.59     & 0.50       & 0          & 0.49     & 0.89       & 0          & 0.90   & 0.46       & 0          &  1  \\
                         \hline
        $D^0 \to K^{*-}$ \cite{Melikhov:2000yu} 
                         & 0.76     & 0.17       & 0          & 0.66     & 0.3        & 0          & 0.49     & 0.67       & 0          & 1.03   & 0.27       & 0          &  1\\ 
                         \hline
        $D^+ \to \rho^0 $ \cite{Melikhov:2000yu} 
                         & 0.66     & 0.36       & 0          & 0.59     & 0.50       & 0          & 0.49     & 0.89       & 0          & 0.90   & 0.46       & 0          &  1/2  \\
                         \hline
        $D^+ \to \omega$ \cite{Melikhov:2000yu} 
                         & 0.66     & 0.36       & 0          & 0.59     & 0.50       & 0          & 0.49     & 0.89       & 0          & 0.90   & 0.46       & 0          &  1/2 \\
                         \hline
        $D^+ \to \overline{K^{*0}}$ \cite{Melikhov:2000yu} 
                         & 0.76     & 0.17       & 0          & 0.66     & 0.3        & 0          & 0.49     & 0.67       & 0          & 1.03   & 0.27       & 0          &  1  \\
                         \hline
      $D_s^+ \to K^{*0}$ \cite{Melikhov:2000yu} 
                         & 0.67     & 0.2        & 0          & 0.57     & 0.29       & 0.42       & 0.42     & 0.58       & 0          & 1.04   & 0.24       & 0          &  1  \\
                         \hline
        $D_s^+ \to \phi$ \cite{Melikhov:2000yu} 
                         & 0.73     & 0.10       & 0          & 0.64     & 0.29       & 0          & 0.47     & 0.63       & 0          & 1.10   & 0.26       & 0          &  1  \\
                         \hline
    $B^+ \to \rho^0$ \cite{Melikhov:2000yu} 
                         & 0.30     & 0.54       & 0          & 0.26     & 0.73       & 0.1        & 0.29     & 1.4        & 0.5        & 0.31   & 0.59       & 0          &  1/2  \\
                         \hline
        $B^+ \to \omega$ \cite{Melikhov:2000yu} 
                         & 0.30     & 0.54       & 0          & 0.26     & 0.54       & 0.1        & 0.24     & 1.40       & 0.50       & 0.31   & 0.59       & 0          &  1/2 \\
                         \hline
$B^+ \to \overline{D^{*0}}$  \cite{Melikhov:2000yu}
                         & 0.69     & 0.58       & 0          & 0.66     & 0.78       & 0          & 0.62     & 1.04       & 0          & 0.76   & 0.57       & 0          &  1 \\
                         \hline
        $B^0 \to \rho^-$ \cite{Melikhov:2000yu} 
                         & 0.30     & 0.54       & 0          & 0.26     & 0.54       & 0.1        & 0.24     & 1.40       & 0.50       & 0.31   & 0.59       & 0          &  1   \\
                         \hline
        $B^0 \to D^{*-}$  \cite{Melikhov:2000yu}
                         & 0.69     & 0.58       & 0          & 0.66     & 0.78       & 0          & 0.62     & 1.04       & 0          & 0.76   & 0.57       & 0          &  1  \\
                         \hline
    $B_s^0 \to K^{* - }$ \cite{Melikhov:2000yu} 
                         & 0.37     & 0.60       & 0.16       & 0.29     & 0.86       & 0.6        & 0.26     & 1.32       & 0.54       & 0.38   & 0.66       & 0.30       &  1 \\
                         \hline
    $B^0_s \to D_s^{*-}$ \cite{Faustov:2012mt} 
                         & 0.67     & 0.35       & 0          & 0.70     & 0.463      & 0          & 0.75     & 1.04       & 0          & 0.95   & 0.372      & 0          &  1  \\
                         \hline
      $B_c^+ \to D^{*0}$ \cite{Ivanov:2000aj} 
                         & 0.56     & 0          & 0          & 0.64     & 0          & 0          & -1.17    & 0          & 0          & 0.98   & 0          & 0          &  1    \\  

    \hline
    \end{tabular}
    \caption{The parameters $f_0(0)$, $\sigma_1$, $\sigma_2$, and $c_V$ which appear in the $P \to V l N$ branching fraction expressions \cref{eq:3bodypseudoscalar,eq:vec form factors1,eq:vec form factors2}.}
    \label{tab:vector form factors}
\end{table}

\begin{table}[bp]
    \centering
    \begin{tabular}{|l|S|S|S|S|S|S|S|S|S|S|D{.}{.}{4}|S|c|}
        \hline
                   & \multicolumn{3}{c|}{$A_0$}  & \multicolumn{3}{c|}{$A_1$} &  \multicolumn{3}{c|}{$A_2$}  & \multicolumn{3}{c|}{$V$}  & \\              
        \hline
        Decay Channel    & \multicolumn{1}{c|}{$f_0(0)$} & \multicolumn{1}{c|}{$\delta$} & \multicolumn{1}{c|}{$m_{\text{fit}}$} & \multicolumn{1}{c|}{$f_0(0)$} & \multicolumn{1}{c|}{$\delta$} & \multicolumn{1}{c|}{$m_{\text{fit}}$} & \multicolumn{1}{c|}{$f_0(0)$} & \multicolumn{1}{c|}{$\delta$} & \multicolumn{1}{c|}{$m_{\text{fit}}$} & \multicolumn{1}{c|}{$f(0)$} & \multicolumn{1}{c|}{$\delta$} & \multicolumn{1}{c|}{$m_{\text{fit}}$} & $c_V$\\
        \hline
     $B_c^+ \to J/ \psi$ \cite{Ivanov:2000aj}
                         & 0.68     & 1.40       & 8.20          & 0.68     & 0.052      & 5.91          & -0.004   & -0.004     & 5.67          & 0.96   & 0.0013     & 5.65          &  1  \\
                         \hline
      $B_c^+ \to B^{*0}$ \cite{Ivanov:2000aj}
                         & -0.27    & 0.13       & 1.86         & 0.6      & -1.07      & 3.44          & 10.8     & -0.09      & 1.73          & 3.27   &-0.052      & 1.76          &  1  \\
                         \hline
    $B_c^+ \to B_s^{*0}$ \cite{Ivanov:2000aj} 
                         & -0.33    & 0.13       & 1.86          & 0.4      & -1.07      & 3.44          & 10.4     & -0.09      & 1.73          & 3.27   &-0.052      & 1.76          &  1  \\
    \hline
    \end{tabular}
    \caption{Parameters $f_0(0)$, $\delta$, $m_{\text{fit}}$, and $c_V$ for certain $B_c \to V l N$ decays appearing in branching fraction expressions \cref{eq:3bodypseudoscalar} and \cref{eq:B_c form factors}.}
    \label{tab:vector form factors B_c}
\end{table}

\subsection{2-Body Tau Decays \boldmath{$\tau \to P N$}}
\label{sbsec:2bdy_tau}

The branching fraction for the $\tau \to P N$ decays are~\cite{Gorbunov:2007ak}
\begin{equation}
\begin{split}
B(\tau \to PN) =& \ \tau_\tau \left|U_\tau\right|^2 \ \frac{G_F^2 m_\tau^3  \left|V_P\right|^2 f_P^2 }{16 \pi} \left[\left(1-\frac{m_N^2}{m_\tau^2}\right)^2-\frac{m_P^2}{m_\tau^2}\left(1+\frac{m_N^2}{m_\tau^2}\right)\right]\\
 &\times \sqrt{\left(1-\frac{\left(m_P-m_N\right)^2}{m_\tau^2}\right)\left(1-\frac{\left(m_P+m_N\right)^2}{m_\tau^2}\right)} \ ,
\end{split}
\end{equation}
where $V_P$ is the CKM matrix element of $P$ (e.g., if $P= \ \pi$, then $V_P= \ V_{ud}$). 
The $\tau$ lepton can also decay into a vector meson.  In this case, the branching fraction is~\cite{Gorbunov:2007ak}
\begin{equation}
\begin{split}
B(\tau \to V N) = & \ \tau_\tau \left|U_\tau\right|^2 \frac{G_F^2  m_\tau^3 \left|V_{V}\right|^2 f_V^2 }{8 \pi }  \left[\left(1-\frac{m_N^2}{m_\tau^2}\right)^2+\frac{m_V^2}{m_\tau^2}\left(1+\frac{m_N^2-2 m_V^2}{m_\tau^2}\right)\right]\\
 &\times \sqrt{\left(1-\frac{\left(m_V-m_N\right)^2}{m_\tau^2}\right)\left(1-\frac{\left(m_V+m_N\right)^2}{m_\tau^2}\right)} \ ,
\end{split}
\end{equation}
where $V_V$ is the CKM matrix element for the corresponding vector meson. Note that in the literature, this branching fraction is often written in terms of the coupling $g_{V}= \ m_V f_V$~\cite{Bondarenko:2018ptm}.

\subsection{3-Body Tau Decays \boldmath{$\tau \to \nu l N, \bar{\nu} l N$}}
\label{sbsec:3bdy_tau}

Lastly, we consider 3-body $\tau$ decays. The differential branching fractions are~\cite{Gorbunov:2007ak}
\begin{equation}
\begin{split}
\frac{d B\left(\tau \to \nu_\tau l_\alpha N\right)}{d E_N} =& \ \tau_\tau  \left|U_\alpha\right|^2  \frac{G_F^2 m_\tau^2}{2 \pi^3}     E_N \left(1-\frac{m_l^2}{m_\tau^2+m_N^2-2 E_N m_\tau}\right) \sqrt{E_N^2-m_N^2}\\
& \times \left(1+\frac{m_N^2-m_l^2}{m_\tau^2}-2 \frac{E_N}{m_\tau}\right) \ ,
\end{split}
\end{equation}
and
\begin{equation}
\begin{split}
\frac{d B \left(\tau \to \bar{\nu}_\alpha l_\alpha N\right)}{d E_N} =& \ \tau_\tau \left|U_\tau\right|^2 \frac{G_F^2 m_\tau^2}{4 \pi^3} \left(1-\frac{m_l^2}{m_\tau^2+m_N^2-2 E_N m_\tau}\right)^2 \sqrt{E_N^2-m_N^2}\\
 &\times\left[\left(m_\tau-E_N\right)\left(1-\frac{m_N^2+m_l^2}{m_\tau^2}\right) \right.\\
 & \left. -\left(1-\frac{m_l^2}{m_\tau^2+m_N^2-2 E_N m_\tau}\right)\left(\frac{\left(m_\tau-E_N\right)^2}{m_\tau}+\frac{E_N^2-m_N^2}{3 m_\tau}\right)\right] \ ,
\end{split}
\end{equation}
where $\alpha \neq \tau$. The bounds of integration are
\begin{equation} 
m_N \leq \ E_N \leq \ \frac{m_{\tau}^2 + m_N^2 -m_{l}^2}{2 m_{\tau}} \ .
\end{equation}

\section{HNL Decay Rates
\label{apx:decay}}

HNL decays are induced by both CC and NC interactions. These interactions are a result of mixing with the SM neutrino gauge eigenstates, which can be seen in the following interacting Lagrangian:
\begin{align}
    -\mathcal{L}^{CC} = & \ \frac{g}{\sqrt{2}}\sum_{\alpha = \ e,\mu,\tau} U^*_\alpha  W_\mu^+  \,\overline{N^c}\gamma^\mu P_L l_\alpha + \text{h.c.} \ , \\
    -\mathcal{L}^{NC} = & \ \frac{g}{2\cos\theta_W}\sum_{\alpha = \  e,\mu,\tau} U^*_\alpha Z_\mu  \overline{N^c}\gamma^\mu P_L \nu_\alpha+ \text{h.c.} \ .
\end{align}

In our analysis of HNL decays, we adopt the convention to Ref.~\cite{Ballett:2019bgd}, where decay rates are summed over all final neutrino mass eigenstates, giving the relationship

\begin{equation}
    \Gamma(N\to \nu X) = \ \sum_{i= \ 1,2,3} \Gamma(N \to \nu_i X) = \  \sum_{\alpha= \ e,\mu,\tau} \Gamma(N \to \nu_\alpha X) + \Gamma(N \to \overline{\nu}_\alpha X) \ .
\end{equation}

The benefit of this choice is that it allows one to remain agnostic about whether light active neutrinos, $\nu_\alpha$ are treated as Dirac or Majorana particles~\cite{Ballett:2019bgd}, a convention choice that leads to non-physical, factor-of-two discrepancies throughout the literature~\cite{Gorbunov:2007ak,Atre:2009rg,Gribanov:2001vv,Helo:2010cw,Bondarenko:2018ptm}.

\subsection{Leptonic Decays}

HNLs decay into purely leptonic final states through the CC and NC processes shown in \cref{fig:feynman_decay}(a)-(c).  The decay widths are~\cite{Coloma:2020lgy}
\begin{align}
    \Gamma ( N \to \nu \, l_\alpha^- l_\alpha^+) = & \ \frac{G_F^2 m_N^5}{96\pi^3}\sum_{\beta = \  e,\mu,\tau} |U_{\beta}|^2 \left[(C_1^l + 2\sin^2\theta_W\delta_{\alpha\beta})f_1(x_\alpha) + (C_2^l+\sin^2\theta_W\delta_{\alpha\beta})f_2(x_\alpha)\right],\\
    \Gamma ( N \to \nu \, l_\alpha^- l_\beta^+) = & \ \frac{G_F^2 m_N^5}{192 \pi^3} \left[|U_{\alpha}|^2 I_1(0,x_\alpha^2,x_\beta^2) + |U_{\beta}|^2 I_1 (0,x_\beta^2,x_\alpha^2) \right] \, \,\,\,\,\,(\alpha\neq\beta)\ ,\\
    \Gamma ( N \to \nu \overline{\nu}\nu) = & \ \sum_{\alpha= \ e,\mu,\tau} \sum_{\beta= \ e,\mu,\tau}\Gamma_{\pm} ( N \to \nu_\alpha \overline{\nu}_\beta \nu_\beta) = \    \frac{G_F^2 m_N^5}{96\pi^3}\left(\sum_{\alpha= \ e,\mu,\tau} |U_{\alpha }|^2\right) \ ,
\end{align}
where $x_i = \ \frac{m_i}{m_N}$, $C^l_1 = (1-4\sin^2\theta_W + 8 \sin^4\theta_W)/4$, and $C^l_2 = (2\sin^4\theta_W-\sin^2\theta_W)/2$. The kinematic functions, $f_1(x)$, $f_2(x)$, and $I_1(x,y,z)$ are given by: 

\begin{align}
    I_1(x, y, z) = & \ 12 \int_{(\sqrt{x}+\sqrt{y})^2}^{(1-\sqrt{z})^2} \frac{d s}{s}\left(s-x-y\right)\left(1+z-s\right) \lambda^{\frac{1}{2}}(1, x, y) \lambda^{\frac{1}{2}}(1,s,z), \\
    f_1(x) = & \ (1-14x^2-2x^4-12x^6)\sqrt{1-4x^2} + 12x^4(x^4-1)L(x), \\ 
     f_1(x) = & \ 4\left[x^2(2+10x^2-12x^4)\sqrt{1-4x^2} + 6x^4(1-2x^2+2x^4)L(x)\right],
\end{align}
where
\begin{align}
    L(x) = & \ \ln\left(\frac{1-3x^2-(1-x^2)\sqrt{1-4x^2}}{x^2(1+\sqrt{1-4x^2})}\right),\\
    \lambda(x,y,z) = & \  x^2 + y^2 + z^2 -2xy - 2yx - 2xz \ . 
\end{align}
For $m_N < \ m_\pi$, the dominant decay is the invisible decay $N\to \nu\overline{\nu}\nu$. However, for $m_N > \ m_\pi$, the 2-body semi-leptonic pion decay, $\nu \pi$, becomes kinematically accessible and begins to dominate the HNL decay width. This can be observed explicitly in \cref{fig:decay br 011,fig:decay br 111}.

\subsection{Hadronic Decays}

\subsubsection{Single Meson Decay}

HNL decays into hadronic final states occur through the NC and CC processes depicted in \cref{fig:feynman_decay} (d)-(e). As mentioned in \cref{sec:decays}, for $m_N < 1.0$ GeV, the total hadronic width is calculated via decay into single meson final states. The formulae for decay of HNLs into single mesons are~\cite{Coloma:2020lgy}  

 \begin{align}
     \Gamma(N \to \nu P^0) =& \ \left(\sum_{\beta= e,\mu,\tau} |U_{\beta }|^2\right) \frac{G_F^2 f_P^2 m_N^3}{16\pi} (1-x_P^2)^2&& [P^0 = \  \pi^0, \eta,\eta\prime] \ ,\\
     \Gamma(N \to l_\alpha^\mp P^\pm) = &\   |U_\alpha|^2 |V_{q\overline{q}}|^2 \frac{G_F^2 f_P^2 m_N^3}{16\pi} \lambda^{\frac{1}{2}}(1,x_P^2,x_\alpha^2)\left[1-x_P^2 -x_\alpha^2(2+x_P^2 - x_\alpha^2)\right] &&[P^\pm = \ \pi^\pm, K^\pm,D^\pm,D_s^\pm] \ ,\\
    \Gamma(N \to \nu V^0) = & \ \left(\sum_{\beta= e,\mu,\tau} |U_{\beta }|^2\right) \frac{G_F^2 f_V^2 \kappa_V^2 m_N^3}{16\pi} (1+2x_V^2) (1-x_V^2)^2&&[V^0 = \ \rho, \omega,\phi] \ ,\\ 
    \Gamma(N \to l_\alpha^\mp V^\pm) = &\ \ |U_\alpha|^2 |V_{q\overline{q}}|^2 \frac{G_F^2 f_V^2 m_N^3}{16\pi}\lambda^{\frac{1}{2}}(1,x_V^2,x_\alpha^2)\left[(1-x_V^2)(1+2x_V^2)+x_\alpha^2(x_V^2 + x_\alpha^2-2)\right] && [V^\pm = \ \rho^\pm, {K^*}^\pm] \ ,
 \end{align}
where $\kappa_\rho = \ 1 - 2\sin^2\theta_W$, $\kappa_\omega = \ -2\sin^2\theta_W/3$, and $\kappa_\phi = \ -\sqrt{2}(1/2 - 2\sin^2\theta_W/3)$ . The meson decay constants $f_{P, V}$ are given in \cref{tab:decay constants}.

\subsubsection{Multi-meson Decays}

In the regime, $m_N > 1.0 $ GeV, the total hadronic width is instead calculated via decays into quarks, since in this region decays into multi-meson states become relevant to the total width. The tree-level NC and CC decays into quarks are given by~\cite{Bondarenko:2018ptm}
\begin{align}
     \Gamma(N\to\nu q\overline{q}) = & \ \left(\sum_{\beta= e,\mu,\tau} |U_{\beta }|^2\right) \frac{G_F^2 m_N^5}{32\pi^3} \left[C_1^q f_1(x_\alpha) + C_2^qf_2(x_\alpha)\right] \ , \\ 
     \Gamma(N\to l_\alpha^- u\overline{d}) = & \ \Gamma(N\to l_\alpha^+ \overline{u}d) = \ |U_\alpha|^2 |V_{ud}|^2 \frac{G_F^2 m_N^5}{64\pi^3} I_1(x_\alpha^2,x_u^2,x_d^2) \ ,
 \end{align}
where the coefficients $C^q_1$ and $C^q_2$ are given by 
\begin{align}
    C_1^{u,c} = & \ \frac{1}{4}\left(1-\frac{8}{3}\sin^2\theta_W + \frac{32}{9}\sin^4\theta_W\right), & C_2^{u,c} = & \ \frac{1}{3}\sin^2\theta_W \left( \frac{4}{3}\sin^2\theta_W - 1\right), \\ 
    C_1^{d,s,b} = & \ \frac{1}{4}\left(1-\frac{4}{3}\sin^2\theta_W + \frac{8}{9}\sin^4\theta_W\right), & C_2^{d,s,b} = & \ \frac{1}{6}\sin^2\theta_W\left(\frac{2}{3}\sin^2\theta_W - 1 \right).
\end{align}
As is discussed in Refs.~\cite{Coloma:2020lgy, Bondarenko:2018ptm}, the total hadronic width can be estimated using this tree-level decay along with a QCD loop correction that accounts for the hadronization process. This correction is estimated from hadronic tau decay.  The loop corrections can be defined as 
\begin{equation}
    1 + \Delta_{\text{QCD}} \equiv \frac{\Gamma(\tau \to \nu_\tau + \text{hadrons})}{\Gamma_\text{tree} (\tau \to \nu_\tau u \bar{d}) 
    +\Gamma_\text{tree} (\tau \to \nu_\tau u \bar{s})} \ ,
\end{equation}\
where the correction up to $\mathcal{O}(\alpha_s^3)$ has been calculated to be 
\begin{equation}
    \Delta_\text{QCD} = \frac{\alpha_s}{\pi} + 5.2\frac{\alpha^2_s}{\pi^2} + 26.4 \frac{\alpha^3_s}{\pi^3} \ .
\end{equation} 
We apply this correction to the NC decays $\nu q\bar{q}$ with $q = u,d,s$, and the CC decays $l u \bar{d}$ and $l u \bar{s}$, where $\alpha_s=\alpha_s(m_N)$. The package \texttt{rundec} is used to compute the $\alpha_s$ running; see Ref.~\cite{Herren:2017osy}. Additionally, for the decay $\nu s s$, a phase space suppression factor, $\sqrt{1-4m_K^2/m_N^4}$, is included to not overestimate this mode's contribution for $m_N < 2m_K$. This same approach is also applied to the decays $\tau ud$ and $ \tau us$, with kinematic thresholds at $m_\tau+2m_\pi$ and $m_\tau + m_\pi + m_K$, respectively.

\subsection{HNL Total Decay Width}
 \label{sec:total width}

The total decay width of the HNL is obtained by summing over all possible final state decays. Since we are considering Majorana HNLs, the CC decays $N\to l_\alpha^+ H^- (\overline{u}d)$, $\nu \,l_\alpha^+ l_\beta^- $ and their charge conjugates are both possible, and, since they have equal decay rates, a factor of 2 must be included when calculating the total rate.  In the region $m_N < 1.0$ GeV, the total decay width is 
 \begin{equation}
 \begin{split}
     \Gamma_N = & \ \Gamma(N\to \nu \overline{\nu}\nu) + \sum_{\alpha =  e,\mu,\tau}\Gamma(N\to\nu \,l_\alpha^+ l_\alpha^-) +\sum_{P^0 =  \pi^0 \eta,\eta^\prime} \Gamma(N \to \nu\,P^0)+ \sum_{V^0 =  \rho^0,\omega,\phi} \Gamma(N \to \nu\,V^0) \\
     & +\sum_{\alpha \neq   \beta =  e,\mu,\tau}2\Gamma(N\to\nu \,l_\alpha^- l_\beta^+)  +   \sum_{\substack{P = \pi,K, D, D_s \\ \alpha =  e,\mu,\tau}} 2\Gamma(N\to l_\alpha^- P^+)+ \sum_{\substack{V =  \rho,K^* \\ \alpha =  e,\mu,\tau}} 2\Gamma(N\to l_\alpha^- V^+) \ . 
 \end{split}
 \end{equation}
For $m_N > 1.0$ GeV, this becomes 
 \begin{equation}
 \begin{split}
     \Gamma_N =  & \ \Gamma(N\to \nu \overline{\nu}\nu) + \sum_{\alpha =  e,\mu,\tau}\Gamma(N\to\nu \,l_\alpha^+ l_\alpha^-) +\sum_{q = u,d,s,c,b} \Gamma(N \to \nu\,q\overline{q})  \\
     & +\sum_{\alpha \neq   \beta =  e,\mu,\tau}2\Gamma(N\to\nu \,l_\alpha^- l_\beta^+)  +   \sum_{\substack{u =  u,c \\ d =  d,s,b}} 2\Gamma(N\to l_\alpha^- u\overline{d}) \ .
 \end{split}
 \end{equation}

\bibliography{hnls}

\end{document}